\documentclass[prx,twocolumn,english,superscriptaddress,floatfix,aps,10pt,nolongbibliography,nofootinbib]{revtex4-2} 
\usepackage{amsmath, amssymb, amscd, amsthm, amsfonts, amscd, dsfont, mathtools, wasysym, amsthm}
\usepackage{graphicx}%
\usepackage{dcolumn}%
\usepackage{bm}
\usepackage{standalone}
\usepackage[caption=false]{subfig}
\usepackage{enumitem}
\usepackage[utf8]{inputenc}
\usepackage{microtype}
\usepackage[plain]{algorithm}
\usepackage{algpseudocode}

\usepackage{bbm}
\usepackage{braket}
\usepackage{comment}
\usepackage{booktabs} %
\usepackage{multirow}
\usepackage{diagbox}

\usepackage{pifont}

\usepackage{refcount}
\newcommand{\tabref}[1]{%
  \hyperref[#1]{Table~\MakeUppercase{\romannumeralstring{\getrefnumber{#1}}}}%
}
\newcommand{\romannumeralstring}[1]{%
  \expandafter\romannumeral\expandafter#1\relax
}

\usepackage{silence}
\newcommand{\ketbra}[2]{| #1 \rangle \langle #2 | }

\let\Re\relax

\DeclareMathOperator{\Re}{Re}

\DeclareMathOperator{\Tr}{Tr}

\newcommand{\dif}{{\rm d}}

\newcommand{\XL}{\overline{X}}
\newcommand{\YL}{\overline{Y}}
\newcommand{\ZL}{\overline{Z}}
\newcommand{\zeroL}{\ket{\overline{0}}}

\newcommand{\oneL}{\ket{\overline{1}}}

\newcommand{\kpsi}{\ket{\psi}}
\newcommand{\kpsis}{\ket{\psi_s}}
\newcommand{\ens}{{\cal E}}
\newcommand{\ensH}{{\cal E}_{\rm H}}
\newcommand{\id}{\mathds{1}}

\newcommand{\calM}{\hat{\mathcal{M}}_s}
\newcommand{\tilM}{\tilde{\mathcal{M}}_s}
\newcommand{\calZ}{\mathcal{Z}_{q,s}}
\newcommand{\initial}{\ket{\phi^{\text{(i)}}}}
\newcommand{\final}{\ket{\phi_q^{\text{(f)}}}}

\usepackage[dvipsnames]{xcolor}

\usepackage{hyperref}
\hypersetup{
    colorlinks = true,
    linkcolor = Blue,%
    citecolor = MidnightBlue,
    urlcolor = Blue}

\let\oldciteauthor\citeauthor
\renewcommand{\citeauthor}[1]{%
  \hypersetup{citecolor=black}%
  \oldciteauthor{#1}%
  \hypersetup{citecolor=MidnightBlue}%
}
\begin{document}

\title{\scalebox{0.95}{Projected logical ensembles in surface codes via the random-matrix theory of quantum dots}}

\def\TCM{TCM Group, Cavendish Laboratory, University of Cambridge, J.\,J.\,Thomson Avenue, Cambridge CB3 0US, UK}
\def\quantinuum{Quantinuum, Terrington House, 13-15 Hills Rd, Cambridge, CB2 1NL, UK}
\def\DAMTP{DAMTP, University of Cambridge, Wilberforce Road, Cambridge, CB3 0WA, UK}
\def\KCL{Department of Physics, King's College London, Strand WC2R 2LS, UK}

\newcommand{\equalcontrib}{\thanks{These authors contributed equally to this work.}}

\author{Mircea Bejan}
\thanks{These authors contributed equally to this work.}
\affiliation{\KCL}
\affiliation{\TCM}

\author{Jan Behrends}
\thanks{These authors contributed equally to this work.}
\affiliation{\TCM}
\affiliation{\quantinuum}

\author{Max McGinley}
\affiliation{\TCM}

\author{Benjamin B\'eri}
\affiliation{\TCM}
\affiliation{\DAMTP}

\date{June 2026}

\begin{abstract}
Measurements underpin active quantum error correction (QEC) and have been recognized as a source of novel measurement-induced many-body phenomena. Here, we study the statistical properties of post-measurement logical states arising in QEC on topological codes subject to deterministic transversal unitary gates. Upon syndrome extraction followed by maximum-likelihood decoding, a Born-weighted ensemble arises which we dub the ``projected logical ensemble'' (PLE). Focusing on surface codes subject to uniform  single-qubit Pauli-$X$ rotations, we characterize the measurement-induced randomness of the PLE. To this end, we show that for a code with a single logical qubit, the PLE is isomorphic to an ensemble of scattering matrices describing mesoscopic quantum dots obtained from a 2D Majorana network model with suitable boundary conditions. We uncover regimes where these quantum dots are chaotic such that their scattering matrices are well-described by random matrix theory. In these regimes,  the PLE approaches a universal ensemble that is maximally random up to symmetry and decoder-induced constraints. The symmetry constraints, set by stabilizer and logical operator weights, realize Altland-Zirnbauer classes D or DIII, which we both illustrate. Our results establish a fundamental connection between emergent universality concepts in mesoscopic physics, quantum many-body systems, and QEC.
\end{abstract}
\maketitle

\section{Introduction}

Quantum error correcting (QEC) codes encode logical quantum information in the state of a large number of physical degrees of freedom, and form the basis of schemes for storing and manipulating quantum information in a way that is protected against hardware-level noise~\cite{shor, CalderbankShor, steane, gottesman1997stabilizer, Aharonov97, Knill_1998,Bravyi1998, Dennis:2002ds,Kitaev:2003jw,gkp, qLDPC,terhal,nielsen2010quantum}. While there exist many different choices of code and various modes of operation, any successful implementation of quantum computation is likely to need some form of QEC for scalability.  

The most prevalent QEC paradigm is %
\textit{active} QEC~\cite{shor, CalderbankShor, steane, gottesman1997stabilizer, Aharonov97, Bravyi1998, Knill_1998,Dennis:2002ds,Kitaev:2003jw,gkp, qLDPC,terhal,nielsen2010quantum, exp1, exp2, Google2023, exp3, exp4, bluvstein2026fault,Quantinuum2026}. Here, ``syndrome'' measurements are performed on the physical degrees of freedom in a targeted way, aiming to detect noise-induced errors that have previously occurred, to then compensate these via a feedback operation. %
For example, in the surface code (and more general topological codes), the encoded states can be interpreted as degenerate ground states of a local many-body Hamiltonian, and the measurements reveal the presence of noise-generated excitations above this ground state~\cite{Bravyi1998,Kitaev:2003jw,Dennis:2002ds}. In addition to their role in stabilizing quantum memories, measurements can also be used to deliberately manipulate the encoded logical information~\cite{Gottesman_1999,PhysRevLett.86.5188,Bravyi_Kitaev_PhysRevA.71.022316,Horsman_2012}, e.g., via lattice surgery~\cite{Horsman_2012}, thereby enabling certain logical gates.

While the usefulness of measurements in QEC has long been appreciated, recently a distinct direction has emerged focusing on the effect of measurements on many-body quantum states relevant to condensed matter~\cite{skinner2019mipt,li2018zeno,li2019mipt,Choi_PhysRevLett.125.030505,gullans2020purification,ippoliti2021mom,fisher2023rqc,Jian:2022jg}. Upon measuring some subset of degrees of freedom, the state of rest of the system will be disturbed in a non-deterministic way due to measurement back-action, thereby inducing a random ensemble of post-measurement states referred to as the \textit{projected ensemble}. Through studying different statistical properties of this ensemble, several novel measurement-induced many-body phenomena have been discovered, including deep thermalization~\cite{Cotler2023, JChoi_2023, HoChoi2022,  Ippoliti2022solvablemodelofdeep, Claeys2022, MaxMichelePRL, Tanmay2023, Ippoliti2023purification, dgge,mark2024maximumentropyprincipledeep, liu2024deepthermalizationcontinuousvariablequantum,  chang2024deepthermalizationchargeconservingquantum, Bejan:2025ch,mixedstateDTKos,mcginley2025scroogeensemblemanybodyquantum,mok2026naturestingyuniversalityscrooge,changCoh}, long-range measurement-induced entanglement~\cite{mie1, mie2, mie3, mie4, mie5, mie6, mie7}, and measurement-induced and measurement-altered criticality~\cite{mic1, mic2, PRXQuantum.4.030317, mic3, mic4,Jian_MIC_PhysRevB.101.104302,Fuji_PhysRevB.102.054302,Turkeshi_PhysRevB.102.014315,Gullans_PhysRevLett.125.070606,negari2026critical}.

The physical interpretation of topological code states as many-body ground states and the importance of measurements in active QEC raise intriguing questions: how do projected ensembles fit within the landscape of QEC?  Conversely, what novel aspects of the projected ensemble—and other measurement-induced phenomena—can be revealed or realized by QEC codes, which are now under intense experimental investigation~\cite{exp1, exp2, Google2023, exp3, exp4, bluvstein2026fault,Quantinuum2026}?

In this work, by leveraging these connections between many-body physics and QEC, we study the statistical properties of post-measurement logical states that arise in surface codes subject to onsite unitary evolution. In QEC terms, we consider deterministic transversal gates, i.e., gates that apply a fixed unitary to each physical qubit. Subsequent syndrome measurements produce random post-measurement states with probabilities determined by Born's rule. These states can each be subsequently brought back to the code space via a maximum-likelihood Pauli recovery. We call the resulting ensemble of logical states the ``projected logical ensemble'' (PLE). We study the statistical properties of the PLE as a function of the distance of the unitary gates from the identity, focusing on codes with a single logical qubit. When the gates are near the identity, typical syndromes are correctable, and thus the recovery operation always brings the system back near the same, deterministic logical state. By contrast, once the gates deviate from identity beyond a threshold, the net process of applying unitaries, measuring syndromes, and applying the correction has a nontrivial, non-deterministic impact on the logical information. Above threshold, states in the PLE become highly random, and we analyze how this distribution depends on the underlying physical parameters.

One of our key contributions is a universal physical picture that captures the properties of the PLE. To formulate this picture, we establish a surprising connection to mesoscopic physics: Conditioned on measurement outcomes, we find that the expectation values of the final logical state can be mapped to scattering amplitudes of a quantum dot constructed from a random single-particle scattering network. The randomness in this system is generated solely by the measurement outcomes, i.e., the Born rule. For the codes studied here, this network and dot belong to Altland-Zirnbauer symmetry classes~\cite{AltlandZirnbauer} DIII or D, as determined by the properties of the code's stabilizers and logical operators.

We identify regimes where this quantum dot is chaotic, which results in scattering amplitudes that are governed by random matrix theory (RMT). Via our mapping, these RMT predictions can be used to characterize the PLE. For class DIII, RMT predicts that the PLE should approximate the Haar ensemble (the maximally random ensemble of pure quantum states), %
which resembles the phenomenon of deep thermalization studied in many-body physics. In class D, the ensemble is instead randomly distributed on a submanifold of states. 

Using direct numerical simulations of the PLE, and by computing properties of the associated network model, we corroborate this picture in a number of different settings. As well as offering a new perspective on the effect of measurements on code states, our connection to mesoscopic physics also provides a new explanation for some of the numerical observations of Refs.~\cite{cheng2025,eckstein2025learningtransitionstopologicalsurface} (albeit in a slightly different setup). Overall, our study highlights the fruitful relationship between QEC codes, mesoscopic physics, and the landscape of measurement-induced phenomena.

\begin{figure*}[htp]
    \centering
    \includegraphics[width=\textwidth]{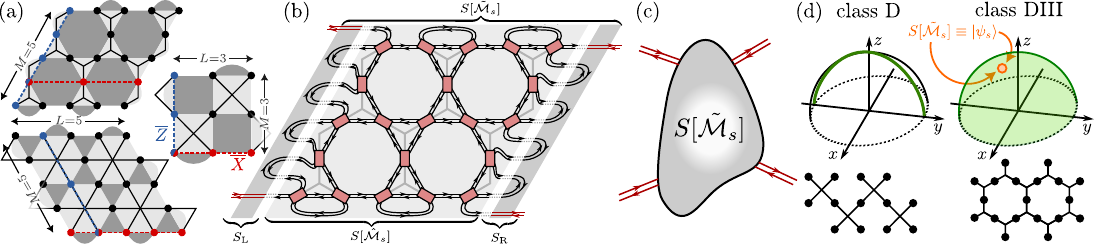}
    \caption{Overview of results. (a): Surface code on honeycomb, square, and triangular lattices with qubits (black disks) on edges (black lines), $S_v^X$ stabilizers (light gray) on vertices, and $S_f^Z$ stabilizers (dark gray) on faces. The logical $\ZL$ (blue) and $\XL$ (red) are also shown. The medial lattice has qubits on its vertices and the faces form a checkerboard pattern of $S_v^X$ and $S_f^Z$ stabilizers.
    (b): The code subject to coherent $X$-rotations and syndrome measurements can be mapped to a quantum circuit (cf.\ Sec.~\ref{sec:network_mapping}), which further maps to a charge-conserving network model~\cite{fvjbbb_prl,Jian:2022jg} with scattering matrix $S[\calM]$ (light gray shaded). Concatenating $S[\calM]$ with scattering matrices $S_\text{L}$ and $S_\text{R}$ (dark gray) describing boundaries with two single-mode contacts each  yields $S[\tilM]$. %
    (c,d): We interpret $S[\tilM]$ as the scattering matrix of a quantum dot and find a mapping between the ensembles of $S[\tilM]$ and the final logical states $\kpsis$, i.e., the PLE. For chaotic quantum dots, $S[\tilM]$ is distributed according to a symmetry-dependent ensemble of random matrices. This random-matrix regime can set in for $\phi>\phi_{\rm th}$ (with $\phi_{\rm th}$ the Pauli-recoverability threshold), and when it does, it implies that the PLE also approaches a universal distribution. %
    For lattices with only even-weight $S^X_v$ (e.g., square or triangular), the quantum dot is in class D; for lattices that include odd-weight $S^X_v$ (e.g., honeycomb) the symmetry class is DIII. Using maximum likelihood decoding, an initial logical $\zeroL$ gets distributed over the highlighted green Bloch line (class D, for odd-weight $\XL$) or hemisphere (class DIII).} %
    \label{fig:overview}
\end{figure*}

\section{Overview}
We begin by summarizing our main results, explaining their relation to previous work, and describing the structure of the paper.
\subsection{Setup and summary of results}

We consider an $n$-qubit system initialized in a code state of a surface code (SC). The code is assumed to have a single logical qubit, but otherwise the geometry of the code can be any planar 2D graph; some examples are shown in Fig.~\ref{fig:overview}(a). To each physical qubit $j \in \{1, \ldots, n\}$, a gate $\exp(i \phi X_j)$ is applied, where the rotation angle $\phi \in [0,\pi/4]$ is deterministic. The code stabilizers are measured, the outcomes of which define a (Born-random) syndrome. Subsequently, a Pauli correction is applied based on a maximum likelihood decoder, which brings the system back into the code space. These final logical states, along with their corresponding Born probabilities, form the ``projected logical ensemble'' (PLE) [Sec.~\ref{sec:LE}], which is the main object of our study.

The combined application of unitary gates and stabilizer measurements on SC states is known to exhibit two phases, corresponding to whether or not the original logical information can be recovered via Pauli correction~\cite{Bravyi:2018ea, Venn:2020ge, fvjbbb_prl, jbfvbb_prr, generic1qCohErr, behrends2025surfacecodepaulichannels,bao2024phasesdecodabilitysurfacecode,yan2026nonlinearsigmamodelsurface,Yang:2026ktu}. Below threshold (i.e., for angles $\phi$ below some critical value $\phi_{\rm th}$), correction succeeds, and the final logical state is with high probability close to the initial logical state. Thus, in this correctable phase, the PLE has a simple structure, with most states close to the same fixed state. Above threshold ($\phi > \phi_{\rm th}$), correction fails and the initial and final logical states differ in a non-deterministic way. Our interest is in the full distribution of final logical states in this non-correctable phase.

As shown in Refs.~\onlinecite{fvjbbb_prl, jbfvbb_prr}, the correctability of logical information in our setting can be analyzed via a mapping to a 2D network model whose geometry is determined by that of the physical qubits [Fig.~\ref{fig:overview}(b); Sec.~\ref{sec:network_mapping}]. This network arises from a transfer matrix that can be interpreted as a (1+1)D nonunitary Gaussian quantum circuit~\cite{jbfvbb_prr,generic1qCohErr, behrends2025surfacecodepaulichannels,bao2024phasesdecodabilitysurfacecode,yan2026nonlinearsigmamodelsurface,Yang:2026ktu}. This reflects a more general principle that relates measurements on isometric tensor network states (a class of states to which the SC state belongs~\cite{isoTNS}) to monitored circuits [Sec.~VII A in Ref.~\onlinecite{mie7}].  An important property of the network model is its symmetry. As we show, depending on the properties of stabilizer generators and logical operators, the network belongs to Altland-Zirnbauer symmetry class D or DIII [Sec.~\ref{sec:symm_lattices}]. The breakdown of correctability at the threshold manifests through a diverging localization length $\xi_{\rm loc}$ in the network. Specifically, for class DIII models, the transition separates an insulating phase for $\phi < \phi_{\rm th}$ from a diffusive metal for $\phi > \phi_{\rm th}$~\cite{yan2026nonlinearsigmamodelsurface,Yang:2026ktu}.

We show that the network model can be used to probe properties of the PLE beyond the question of correctability. In particular, we derive an explicit mapping between the states in the PLE and the scattering matrix $S$ of the network model with suitable boundary conditions [Sec.~\ref{sec:quantum_dot}]. 
The mapping arises from the concatenation of the scattering network and that of boundaries with single-mode contacts, thus leaving 4 incoming and 4 outgoing Majorana modes, see Fig.~\ref{fig:overview}(b). This turns the network into a quantum dot of linear size $L$ proportional to the code distance. 
Each state in the PLE maps---through a fixed isomorphism---to $S$, the dot's scattering matrix. 
Naturally, $S$ inherits the symmetries of the network model. More significantly, the distribution of $S$, which has been extensively studied in mesoscopic physics~\cite{Beenakker:1997gz,AlhassidRMP}, including in classes D and DIII~\cite{AltlandZirnbauer, dahlhaus_prb_2010}, dictates the statistical properties of the PLE. This correspondence is the basis of the physical picture announced in the Introduction.\\

A dot with $L\ll\xi_{\rm loc}$, and $L$ much larger than the mean free path, is a diffusive metallic grain. 
In our setting, the system is in this regime for  $\phi > \phi_{\rm th}$, both for class DIII where it belongs to a metallic phase~\cite{yan2026nonlinearsigmamodelsurface,Yang:2026ktu} and class D where above the numerically observed $\phi_{\rm th}$ the behavior is consistent with such a metallic-grain regime~\cite{fvjbbb_prl,eckstein2025learningtransitionstopologicalsurface}.  
For such diffusive metallic grains, once the escape time much exceeds a timescale inversely proportional to the finite-size scaling of the bulk conductivity $g(L)$~\cite{altshuler1986repulsion,efetov1983supersymmetry,Fyodorov_95_PhysRevB.51.13403,efetov1999supersymmetry,mirlin2000statistics,bocquet2000disordered}, 
transport through the quantum dot is expected to become chaotic, i.e., the dot to act like a random scatterer with no spatial locality [Fig.~\ref{fig:overview}(c)]. In this case, the distribution of $S$ is described by random matrix theory~\cite{Beenakker:1997gz,AlhassidRMP}. Via our mapping, this yields the prediction that the PLE should approach, in a manner controlled by $g(L)$, a maximally random ensemble, subject to symmetry and decoding-induced restrictions [Fig.~\ref{fig:overview}(d)].  For class DIII this is the uniform (Haar) ensemble over all logical states on the upper Bloch hemisphere; for class D, the  distribution is uniform over the upper meridian of the Bloch sphere (in both cases we took $\zeroL$ as the initial logical state). %
We corroborate these predictions with direct numerical simulations of the PLE, alongside simulations of the network models [Sec.~\ref{sec:qdotNum},~\ref{sec:emergentLHE}, App.~\ref{app:TwirlNumerics},~\ref{app:classD}]. 

\subsection{Relation to previous work} \label{sec:relation}
Our work fits into a growing body of literature on QEC codes under coherent noise~\cite{Wallman:2014hs,Wallman:2015cd,Wallman:2016be,DarmawanTensor,Bravyi:2018ea,Greenbaum:2018ce,Gottesman2019,Iverson:2020fe,Beale_PhysRevLett.121.190501,Venn:2020ge, fvjbbb_prl, jbfvbb_prr, generic1qCohErr, behrends2025surfacecodepaulichannels,bao2024phasesdecodabilitysurfacecode,Moon_hchr-rqq9,mie2,Darmawan2025_r2dc-qcrx,Turkeshi_PhysRevLett.132.140401,Lavasani_PhysRevResearch.7.023166,Pato_PhysRevA.111.032424,Huang2025,Nobu_transversal_2025,Sierant_TMPT,wang2025decoherenceinducedselfdualcriticalitytopological,Leblond_ktb3-gcxr,Metodi_2026,cheng2025,eckstein2025learningtransitionstopologicalsurface,Eckstein_teleport_PRXQuantum.5.040313,Putz_qubitloss2025,Putz_learningPRL_4dwm-kn11,yan2026nonlinearsigmamodelsurface,Yang:2026ktu}, 
and is related to recent studies of SCs under coherent rotations and local measurements that discuss state ensembles, thermodynamic phases, or symmetry classes in related settings~\cite{cheng2025,eckstein2025learningtransitionstopologicalsurface,yan2026nonlinearsigmamodelsurface,Yang:2026ktu,Eckstein_teleport_PRXQuantum.5.040313,Putz_qubitloss2025,Putz_learningPRL_4dwm-kn11}. We next describe the relationship between our results and these works.

In Ref.~\cite{cheng2025}, \citeauthor{cheng2025} study Haar-random local unitary rotations on a random subset of physical qubits of an SC with even-weight stabilizers, followed by syndrome measurement and error correction. Under some conditions, this induces unitary logical designs. By contrast, we rotate each qubit by a fixed deterministic single-axis unitary, thus focusing on the Born rule as the only ``source'' of randomness. Furthermore, in our settings, a SC with odd-weight stabilizers is essential for approaching a maximally random PLE. Our quantum dot mapping provides novel physical insights into this emergent universality.

In Ref.~\cite{eckstein2025learningtransitionstopologicalsurface}, \citeauthor{eckstein2025learningtransitionstopologicalsurface} study a state ensemble obtained by entangling a reference qubit with a SC logical qubit, uniformly rotating each SC physical qubit  about a single axis, then measuring each physical qubit along another axis. The SC has even-weight stabilizers. This procedure consumes the SC, and leaves a logical qubit snapshot in the reference. Ref.~\cite{eckstein2025learningtransitionstopologicalsurface} finds numerical results consistent with this logical ensemble approaching Haar distribution in certain regimes they identify as a class DIII Majorana metal. They also identify a ``ring distribution'' analogous to that shown in Fig.~\ref{fig:overview}(d). Refs.~\cite{Eckstein_teleport_PRXQuantum.5.040313,Putz_qubitloss2025,Putz_learningPRL_4dwm-kn11} show how boundary conditions in a dual statistical mechanical description encode logical states. 
By contrast, we measure stabilizers and apply a recovery operation, thereby preserving a SC. We also find class DIII metallic network models, but our construction hinges on the SC having odd-weight stabilizers. Furthermore, we provide a theoretical picture for the emergent universal PLE by mapping it to the RMT of quantum dot scattering.

In Ref.~\cite{yan2026nonlinearsigmamodelsurface}, which appeared during the completion of this manuscript, \citeauthor{yan2026nonlinearsigmamodelsurface} consider a decoding problem for the SC under single-qubit coherent errors. For a bipartite SC lattice and i.i.d. Gaussian rotation angles $\phi$ near $\pi/4$, they derive a class D non-linear sigma model (NL$\sigma$M) as an effective model of the decoding problem, and find an insulating phase under MLD. They argue that for uniform $\phi$ this phase persists, albeit likely with $\xi_\text{loc}\gg L$ for numerically accessible system sizes $L$. For systems where syndromes reside on non-bipartite lattices, they note that a class DIII NL$\sigma$M can arise, and suggest that these can host a thermal metal phase. We consider SC on distinct lattices with uniform $\phi$, which we show are described by a class D or DIII network models. Focusing on the honeycomb SC, our numerics suggest that its disordered network exhibits a class DIII Majorana metal whose transport properties are consistent with a class DIII NL$\sigma$M in the Born-rule replica limit.

In Ref.~\cite{Yang:2026ktu}, which also appeared during the completion of this manuscript, \citeauthor{Yang:2026ktu} study decodability phase transitions for the SC on lattices with different stabilizer-weight parities under single-qubit coherent rotations. They map this problem to (1+1)D nonunitary Gaussian fermion circuits in classes D and DIII, as well as corresponding NL$\sigma$Ms, and identify decodability transitions with entanglement transitions. (For entanglement transition perspectives on decodability, see also Refs.~\cite{jbfvbb_prr,generic1qCohErr,behrends2025surfacecodepaulichannels,bao2024phasesdecodabilitysurfacecode,yan2026nonlinearsigmamodelsurface,cheng2025}.) They find thermodynamic phases consistent with those of \citeauthor{yan2026nonlinearsigmamodelsurface}~\cite{yan2026nonlinearsigmamodelsurface}, and new decodability transitions for non-uniform $\phi$.  We consider the SC on similar lattices with uniform rotations, which we map to class D and DIII nonunitary circuits and to Majorana scattering networks. We identify a similar QEC transition for the honeycomb lattice, and find a class DIII metallic phase above threshold.
Our results are consistent with Refs.~\cite{yan2026nonlinearsigmamodelsurface,Yang:2026ktu}. However, our main focus is the PLE, for which we provide a unified physical picture through a quantum dot mapping. 

\subsection{Outline}
The remainder of this work is organized as follows. In Sec.~\ref{sec:LESC}, we present our SC setup and introduce the PLE. 
In Sec.~\ref{sec:network_mapping}, we outline the mapping to (1+1)D Gaussian quantum circuits, and how the final logical state is embedded in the final states of such a circuit.
In Sec.~\ref{sec:quantum_dot}, we discuss the associated 2D scattering network model, explain the quantum dot mapping, and how the quantum dot scattering matrices are related to the final logical states.
In Sec.~\ref{sec:BornNumerics}, we provide numerical evidence for an emergent chaotic class DIII quantum dot and the emergent PLE. The Appendices contain details of our derivations and additional numerics, including numerics consistent with a chaotic class D dot [App.~\ref{app:classD}].

\section{Projected logical ensemble of the surface code under coherent rotations} \label{sec:LESC}

In this Section, we introduce the SC protocol and ensemble of logical states studied in this work.

\subsection{Surface code setup} \label{sec:SCsetup}
The surface code~\cite{Bravyi1998,Kitaev:2003jw,Dennis:2002ds} is an experimentally relevant~\cite{exp1, exp2, Google2023, exp3, exp4, bluvstein2026fault,Quantinuum2026} quantum Calderbank-Shor-Steane stabilizer code~\cite{CalderbankShor,steane}.
Here we describe its construction on a two-dimensional (2D) planar lattice, with $n$ qubits on its edges, $X$-type stabilizers $S_v^X = \prod_{j \in v} X_j$ on vertices, and $Z$-type stabilizers $S_f^Z = \prod_{j \in f} Z_j$ on faces, where $j\in v$ and $j\in f$ denote the edges $j$ neighboring the vertex $v$ and enclosing the face $f$, respectively.
Since the $S_v^X$ stabilizers are defined on the vertices of a lattice, the $X$-type stabilizers form the direct lattice, the $Z$-stabilizers the dual lattice, and the physical qubits form an auxiliary lattice, the medial lattice.
We focus on the case where one logical qubit is encoded in the SC via the logical $\XL = \prod_{j \in \gamma} X_j$ and $\ZL = \prod_{j\in \gamma'} Z_j$. We choose a minimum-weight representation of the logical operators and consider geometries where $\XL$ has support on qubits on a string from the left to the right of the lattice, and $\ZL$ from top to bottom [Fig.~\ref{fig:overview}(a)].

While often defined on square lattices, the SC can also be defined on other lattices, with benefits for biased noise~\cite{Fuji:2012km,Roethlisberger:2012fh,Venn:2020ge}.
A primary example used in this work is the SC on a honeycomb lattice, as shown in Fig.~\ref{fig:overview}(a).

We now describe the effect of single-qubit coherent rotations followed by stabilizer measurements and Pauli corrections~\cite{Bravyi:2018ea,Venn:2020ge}.
For concreteness, we focus on rotations about the $X$-axis. (Rotations about $Z$ can also be described in the same framework, but the relevant lattice then is that of $Z$-type stabilizers, i.e., the dual lattice~\cite{Venn:2020ge}.) Although we shall be more interested in preparing specific initial logical states ($\ket{\overline{0}}$ or $\ket{\overline{1}}$ for $X$ rotations), it is instructive to specialize to these only later. %

Consider the application of $U=\prod_j \exp (i \phi X_j)$ to an initial logical state $\rho = \ketbra{\psi}{\psi}$, followed by a noiseless syndrome measurement that yields the syndrome $s$; hence, the $X$ syndrome is always trivial (i.e., all $S_v^X$ measurements return $+1$) and only the $Z$ syndrome is nontrivial.
To return to the logical subspace after a syndrome measurement, we apply a Pauli correction $C_s = \prod_j X_j^{(1-\eta_j)/2}$ (with $\eta_j=\pm 1$), which is a string of $X_j$ operators that connects $S_f^Z=-1$ syndromes, either in pairs or to a condensing boundary. The correction $C_s$ shall be chosen using the maximum-likelihood decoder~\cite{Venn:2020ge,fvjbbb_prl}, to be specified later.

For a syndrome $s$, there are two inequivalent correction operations, $C_s$ and $C_s \XL$. $C_s$ can be deformed by $X$ stabilizer configurations as $C_s \mapsto C_s' = C_s \prod_v (S_v^X)^{n_v}$ with some configuration $\{ n_v = 0,1 \}$; since all stabilizers commute, such a deformed $C_s'$ corresponds to the same syndrome.
$C_s$ can also be deformed by a logical $\XL$ operator, $C_s \mapsto C_s \XL$.
Again, since the final state after the correction operation is within the logical subspace, the deformation by an $X$ stabilizer configuration has no effect, as any logical state is in the $+1$ eigenspace of all $S_v^X$.
However, deforming $C_s$ by $\XL$ changes the final state by a logical $\XL$, thus making $C_s$ and $C_s \XL$ inequivalent.

The application of $U$ to an initial state $\rho$, followed by a projection onto the syndrome $s$ and the Pauli correction $C_s$ can be described by the effective logical channel~\cite{Bravyi:2018ea,Venn:2020ge}
\begin{align}
\rho \mapsto \rho_s = 
\frac{C_s \Pi_s U \rho U^\dagger \Pi_s C_s}{p(s\vert \rho)} =\frac{D_s \rho D_s^\dagger}{\Tr [ D_s \rho D_s^\dagger]},
\end{align}
where $\Pi_s$ is the projection onto syndrome $s$ which occurs with probability $p(s\vert \rho) = \Tr [ \Pi_s U \rho U^\dagger \Pi_s] = \Tr [ D_s \rho D_s^\dagger]$, and where we defined the operator $D_s \coloneqq \Pi_0 C_s U \Pi_0$ via the projector $\Pi_0$ onto the logical subspace. Note that $\Pi_s = C_s \Pi_0 C_s$ because $C_s$ maps states from the syndrome $s$ subspace to the logical subspace~\cite{Bravyi:2018ea,Venn:2020ge}. Since we consider only $X$ rotations and accordingly only $X$ corrections, $D_s$ is diagonal in the $X$ basis and we thus have~\cite{Bravyi:2018ea,Venn:2020ge}
\begin{align}
    D_s = (\mathcal{Z}_{0,s} \overline{\id} + \mathcal{Z}_{1,s} \XL) \Pi_0, \label{eq:Ds}
\end{align}with complex coefficients $\mathcal{Z}_{q,s}$.
While the probability $p(s\vert\rho)$ to measure $s$ generally depends on the initial state, it will be useful
to consider its average over the Bloch sphere $p(s) = \int_\Omega \dif \rho\  p(s|\rho) p(\rho) = \sum_q |\calZ|^2$~\cite{Venn:2020ge}.

A key question of this work concerns the statistical properties of this effective logical operation $D_s/\sqrt{p(s|\rho)}$. Specifically, we are interested in the distribution of final logical states $\ket{\psi_s} = D_s\kpsi / \sqrt{p(s|\rho)}$ over the logical Bloch sphere. To draw conclusions about these effective logical operations, we choose the convenient parametrization in terms of two angles $\theta_s \in [0, \pi]$ and $\varphi_s \in [0, 2\pi)$,
\begin{align}
 \mathcal{Z}_{0,s} &= \sqrt{p(s)} e^{i \zeta_s} \cos (\theta_s/2) \label{eq:Z0param}, \\
 \mathcal{Z}_{1,s} &= \sqrt{p(s)} e^{i (\zeta_s+\varphi_s)} \sin (\theta_s/2) \label{eq:Z1param}
\end{align}
such that
\begin{align}
 D_s = \sqrt{p(s)} \left[  \cos (\theta_s/2) \overline{\id} + e^{i \varphi_s} \sin (\theta_s/2) \XL \right] \Pi_0, 
\end{align}
where we ignored the overall inconsequential phase $\zeta_s$.

In the remainder of this work, we consider initial logical states $\kpsi$ that are eigenstates of the logical $\ZL$ operator; that is, we can take $\kpsi = \ket{\overline{0}}$ or $\kpsi =\ket{\overline{1}}$, but focus on $\kpsi = \ket{\overline{0}}$ for clarity. Using this initial state, we can extract the angles of the effective logical operation $\propto D_s$ from the final logical state
\begin{align}
    \ket{\psi_s} =  \cos(\theta_s/2) \zeroL + e^{i \varphi_s} \sin(\theta_s/2) \oneL. \label{eq:psi_s}
\end{align}Note that $p(s) = p(s| \rho =  \ketbra{\overline{0}}{\overline{0}})$~\cite{Venn:2020ge}.

\subsection{Projected logical ensemble} \label{sec:LE}

To characterize the statistical properties of the effective logical operations [Eq.~\eqref{eq:Ds}], we collect the final logical states from Eq.~\eqref{eq:psi_s} in the \textit{projected logical ensemble} (PLE) %
\begin{align}
    {\cal E} \coloneqq \{ p(s|\rho), \ket{\psi_s} \}, \label{eq:LE}
\end{align}with the Born probabilities $p(s|\rho)$ with $\rho = \ketbra{\overline{0}}{\overline{0}}$, and states labeled by the syndrome $s$. 
$\cal E$ is a discrete distribution of points on the Bloch sphere, and we are interested in their statistical properties.

We can gain some intuition by viewing our setup as a QEC problem, where $e^{i\phi X_j}$ act as coherent errors and $\ket{\psi_s}$ are post-QEC states. Below the code's decoder-dependent error threshold, the correction succeeds with high probability; hence, we expect $\cal E$ to be tightly concentrated (increasingly so with increasing $L$) on the Bloch sphere around the initial state $\kpsi$. Conversely, above the error threshold, the correction fails, and the states in $\ens$ may be anywhere on the Bloch sphere, perhaps even uniformly distributed.

In this work, we are interested in the randomness of the PLE $\ens$ in the above threshold regime. The states in the PLE are constrained to lie within a subspace of the Bloch sphere. Before briefly discussing two such constraints (see also App.~\ref{app:constraints}), let us note that there exists a universal ensemble that is maximally random up to constraints, which shall serve as a reference for the maximum randomness that $\ens$ may achieve.
This is the constrained Haar ensemble $\ensH^*$~\cite{harrow2013churchsymmetricsubspace}, which is a continuous, uniform distribution of pure states over a constraint-dependent Bloch-sphere subspace. As we find in Sec.~\ref{sec:emergentLHE}, the numerical data suggests that $\ens$ can approach $\ensH^*$; we discuss how to quantify this approach in Sec.~\ref{sec:emergentLHE} and App.~\ref{app:LErandomness}.

First, the decoder with which we choose a recovery operation from the two inequivalent classes can constrain $\cal E$. Concretely, for the maximum likelihood decoder (MLD) used in this work, the states $\ket{\psi_s}$ may only lie within the upper Bloch hemisphere (given the initial state $\zeroL$ at the North Pole). Denoting two inequivalent corrections by $C_{q,s} = \XL^q C_s$ with $q=0,1$ and the probability of each class by $p_{q,s} = \vert \braket{\psi |\XL^q C_s \Pi_s U | \psi} \vert^2 = \vert {\cal Z}_{q,s} \vert^2$, the MLD selects the class which results in $\kpsis$ closest to $\kpsi$ by picking $C_{q^*, s}$ with $q^* = \arg \max_q p_{q,s}.$ To simplify notations, we use that for a given $s$ we have a given pure state $\Pi_s U \ket{\psi}$ and hence define $C_s$ such that $q^*=0$.
Therefore, for the MLD, $\vert {\cal Z}_{0,s} \vert^2 \geq \vert {\cal Z}_{1,s} \vert^2$ by definition, implying that $\ket{\psi_s}$ must lie in the northern Bloch hemisphere.%

Second, the parity of the weights of $S^X_v$ stabilizers can induce a fixed phase relation between the amplitudes ${\cal Z}_{0,s}$ and ${\cal Z}_{1,s}$~\cite{Bravyi:2018ea, Venn:2020ge}. While we review these considerations in more detail in App.~\ref{app:constraints}, we note that the key insight comes from expanding $U$ and $D_s$ as sums of $X$-Pauli strings, e.g., $U = \sum_{\{ x_j \}} (\cos \phi)^N ( i \sin \phi)^{N-|x|}\prod_j X^{x_j}_{j}$ with $x_j = 0,1$ and Hamming weight $|x|$, and then identifying which $X$-strings contribute to each correction class. One finds that, for codes with even-weight $S^X_v$ stabilizers [such as the square-lattice SC, cf. Fig.~\ref{fig:overview}(a)], there is a fixed phase relation $\arg(\mathcal{Z}_{0,s}/\mathcal{Z}_{1,s}) = \varphi_s \in  \{\varphi, \varphi+ \pi\} \ \forall s$ with $\varphi = 0$ for an even-weight $\XL$, and $\varphi = \pi/2$ for an odd-weight $\XL$; this implies that the logical states $\ket{\psi_s}$ may only lie on a one-dimensional circle subspace of the Bloch sphere. Conversely, for codes with some odd-weight $S^X_v$ stabilizers [including the honeycomb SC, cf. Fig.~\ref{fig:overview}(a)], the relative phase $\varphi_s$ between ${\cal Z}_{0,s}$ and ${\cal Z}_{1,s}$ can take an arbitrary value; thus, in this case, the stabilizer weights do not constrain the PLE.

\section{Quantum circuit} \label{sec:network_mapping}

Here, we relate the amplitudes $\calZ$ [Eq.~\eqref{eq:Ds}] to (1+1)D quantum circuits and discuss how a final 2-qubit projected state of such a circuit encodes the final logical state~[Fig.~\ref{fig:transfer_mapping}].

\subsection{From surface code to (1+1)D quantum circuit}
\label{sec:SC_to_circuit}

The coefficients $\calZ$ [Eq.~\eqref{eq:Ds}] can be efficiently computed using a Gaussian tensor network (TN), which can also be used to efficiently sample syndromes~\cite{Bravyi:2018ea}.
For completeness, in this section we describe the TN contraction and error-string sampling, largely following Ref.~\cite{fvjbbb_prl}.
The starting point is the partition function of a random-bond Ising model (RBIM)
\begin{align}
 \hspace{-0.8em} \calZ &= \sum_{\{\sigma_v\}} e^{n K + H_{q,s} (\{\sigma_v\})}, \   
 H_{q,s} = J \sum_{\langle v,v'\rangle} \eta_{vv'}^{(q,s)} \sigma_v \sigma_{v'}, \label{eq:partfn}
\end{align}
with complex couplings~\cite{fvjbbb_prl}
\begin{align}
  K = \frac{\ln (\cos \phi \sin \phi)}{2} + \frac{i\pi}{4} , & &
  J =-\frac{\ln (\tan \phi )}{2}  - \frac{i\pi}{4} 
\label{eq:rbim_coefficients}.
\end{align}
Ising variables $\sigma_v = \pm1$ on the vertices of the graph with vertices at $S^X_v$ locations and edges on qubits, and $n$ is the total number of qubits. The signs $\eta_{vv'}^{(q,s)} = \pm 1$ can be sampled from a quantum circuit~\cite{Bravyi:2018ea,Venn:2020ge}; in App.~\ref{app:sampling}, we describe a slight variation of the sampling algorithm introduced in Ref.~\onlinecite{Bravyi:2018ea}.
The form of the partition function arises from the expansion of the product $C_s U$ into Pauli strings: these Pauli string can contribute to $D_s$ [Eq.~\ref{eq:Ds}] only when they form products of stabilizers (contributing to $\mathcal{Z}_{0,s}$) or products of stabilizers times $\XL$ (contributing to $\mathcal{Z}_{1,s}$).
The sum over all possible stabilizer configurations can be expressed as a sum over spin configuration, which, using the right choice of the coefficients, gives the partition function~\eqref{eq:partfn}; cf.\ Ref.~\onlinecite{fvjbbb_prl} for more details.

The partition function $\calZ$ equals the contraction of a TN consisting of two-leg tensors $\hat{T}_\eta|_{\sigma\sigma'} =  e^{K +\eta \sigma\sigma'J}$, i.e.,
\begin{align}
 \hat{T}_1 = \begin{pmatrix} \cos \phi & i \sin \phi \\ i \sin \phi & \cos \phi \end{pmatrix} , & &
 \hat{T}_{-1} = \begin{pmatrix} i \sin \phi & \cos \phi \\ \cos \phi &  i \sin \phi \end{pmatrix}
\end{align}
on all qubit sites, i.e., on the edges of the graph.
The two legs of $\hat{T}_\eta$ connect to $S_v^X$ stabilizers on the vertices.
Contracting all legs emanating from a vertex thus corresponds to summing over the two possible stabilizer configurations.

\begin{figure}
    \centering
    \includegraphics[scale=1]{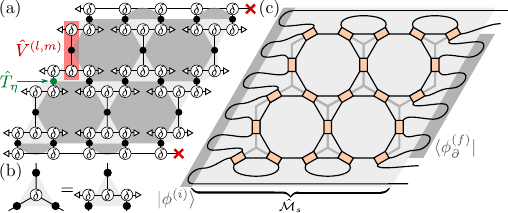}
    \caption{(a): Mapping the sum over stabilizer configurations to a quantum circuit for the honeycomb surface code~\cite{Dennis:2002ds,fvjbbb_prl}.
    Each $S_v^X$ stabilizer is replaced by a Kronecker $\delta$ tensor that can be split up into several $\delta$ tensors (white disks), as shown in panel (b), to ensure a brickwork-structured quantum circuit.
    Combining the tensor $\hat{T}_\eta$ (black disks, one example highlighted in green) with neighboring $\delta$ tensors yields the nonunitary gate $\hat{V}^{(l,m)}$ (red box). The white triangles denote $\sqrt{2}\ket{+}$ states necessary to ensure summing over all stabilizer configurations. Red crosses denote uncontracted legs, resulting in the final state $\ket{\phi_s}$ from Eq.~\eqref{eq:partial_contr}, which encodes the logical state $\kpsis$. (c): The resulting quantum circuit is Gaussian. After mapping to Majorana operators (black wires), $\hat{V}^{(l,m)}$ acts on two neighboring Majorana operators (ocher boxes) that define a quantum circuit $\calM$ (light gray) acting on an initial Gaussian state $\ket{\phi^{(i)}}$ (dark gray). After partial contraction with $\bra{\phi^{(f)}_\partial}$ (dark gray), the four dangling Majorana modes correspond to $\ket{\phi_s}$ (see also Ref.~\cite{Bravyi:2018ea} for a TN perspective).} %
    \label{fig:transfer_mapping}
\end{figure}

We now describe the tensor network as a nonunitary $(1+1)$D quantum circuit~\cite{Bravyi:2018ea,jbfvbb_prr}, i.e., we contract it from left to right.
In Fig.~\ref{fig:transfer_mapping}(a), we show the TN of a honeycomb SC of dimension $(L,M)$ to visualize the network and quantum circuit.
Here we use $(L,M)$ as the lattice dimensions, which do not necessarily equal the code distance; for the honeycomb lattice, while $d_Z = M$ equals the minimum weight of $\ZL$, $d_X = (L+1)/2$ is smaller, as illustrated in Fig.~\ref{fig:overview}(a).
The (unnormalized) initial state of the quantum circuit $\initial = \otimes_{m=1}^{M+1} \sqrt{2} \ket{+}_m$ (not to be confused with $\kpsi$) encodes the boundaries while ensuring the corresponding stabilizer configurations are summed over~\cite{fvjbbb_prl};
note that we inserted two ancilla qubits, $1$ and $M+1$, for later convenience (these will encode $\ket{\psi_s}$, see Sec.~\ref{sec:embedLogical}).
All tensors $\hat{T}_\eta$ are located on vertical bonds, and contracting the index to the left of $\hat{T}_\eta$ in the $l^\text{th}$ layer %
at the $m^\text{th}$ vertical site %
corresponds to acting with the operator $\hat{V}^{(l,m)} = e^{K + \eta^{(l,m)} J \tau_{m}^z \tau_{m+1}^z}$, where the Pauli matrices $\tau_m^\mu$ act in the transfer matrix space of the evolved 1D state; note that $\hat{V}^{(l,m)}$ is not unitary.
For the honeycomb lattice, the even and odd layers contain a different number of qubits and hence a different number of individual $\hat{V}^{(l,m)}$ gates; for odd $l$, $\hat{V}^{(l)} = \prod_{m=1}^{M} \hat{V}^{(l,m)}$ and for even $l$, $\hat{V}^{(l)} = \prod_{m=1}^{(M+1)/2} \hat{V}^{(l,2m-1)}$.
After applying a full layer $\hat{V}^{(l)}$, we apply a layer of horizontal gates that also depends on the choice of the lattice.
For the honeycomb lattice, the horizontal layer consists of alternating identity gates and projections on the $\ket{+}_m$ state times a normalization; after odd-$l$ $\hat{V}^{(l)}$ layers, we apply $\hat{H} = 2^{(M-1)/2}\prod_{m=1}^{(M-1)/2}  \ketbra{+}{+}_{2m}$, and after even-$l$ layers, $\hat{H}' = 2^{(M-1)/2}\prod_{m=1}^{(M-1)/2}  \ketbra{+}{+}_{2m+1}$.
The product of all layers defines the nonunitary quantum circuit~\cite{fvjbbb_prl}
\begin{align}
 \calM = \hat{V}^{(L)} \cdots  \hat{H}  \hat{V}^{(3)} \hat{H}'  \hat{V}^{(2)} \hat{H} \hat{V}^{(1)},
\end{align}
which is a central object of this study.

Contracting the circuit $\calM$ with the initial state $\initial$ and one of the unnormalized final states
\begin{align} \final = \begin{cases} \sqrt{2}^{M-1} \ket{0 + + \cdots + 0}, & q= 0, \\
\sqrt{2}^{M-1} \ket{0 + + \cdots + 1}, & q= 1, 
\end{cases}
 \label{eq:final_states}
\end{align}
yields the coefficient
\begin{align}
 \calZ = \braket{\phi^{\text{(f)}}_q| \calM | \phi^\text{(i)} } .
 \label{eq:contraction}
\end{align}
The states of the first and last qubits in Eq.~\eqref{eq:final_states} are chosen to fix the couplings on the top and bottom rows of qubits; choosing $\ket{0}$ fixes them to the configuration encoded in $\eta^{(l,m)}$, while choosing $\ket{1}$ swaps all bonds in the corresponding row, $\eta^{(l,m)} \to -\eta^{(l,m)}$, which in turn corresponds to inserting a logical $\XL$ that would swap the same bonds. Since inserting the same logical operator twice cancels it, we could have equivalently chosen $\ket{\tilde{\phi}^\text{(f)}_{q=0}} = \sqrt{2}^{M-1} \ket{1 + \cdots+ 1}$ %
and  $\ket{\tilde{\phi}^\text{(f)}_{q=1}} = \sqrt{2}^{M-1} \ket{1 + \cdots + 0}$.%

\subsection{Gaussian (1+1)D quantum circuit}
\label{sec:GaussianCirc}

We next show the circuit $\calM$ is a fermionic Gaussian circuit following Ref.~\cite{Bravyi:2018ea,Venn:2020ge,fvjbbb_prl}, and hence, we can efficiently evaluate it classically, i.e., in polynomial time~\cite{Bravyi:2005jh}.
To this end, we map the operators $\tau_m^\mu$ acting in transfer matrix space to Majorana operators using the Jordan Wigner transformation $\gamma_{2m-1} = (\prod_{k<m} \tau^x_k) \tau^z_m$ and $\gamma_{2m} = (\prod_{k<m} \tau^x_k) \tau^y_m$, such that $i\gamma_{2m-1}\gamma_{2m} = \tau^x_m$ and $i \gamma_{2m}\gamma_{2m+1} = \tau^z_m \tau^z_{m+1}$.
Hence, the gates on vertical bonds are $\hat{V}^{(l,m)} = e^{K + i \eta^{(l,m)} J \gamma_{2m}\gamma_{2m+1}}$, and the $\ket{+}_m$ states correspond to the empty $\ket{0^\text{F}}_m$ with $i\gamma_{2m-1} \gamma_{2m} \ket{0^\text{F}}_m = \ket{0^\text{F}}_m$.
The projectors onto $\ket{+}_m$ thus transform to $\ketbra{0^\text{F}}{0^\text{F}}_m = (1+ i\gamma_{2m-1}\gamma_{2m})/2$, %
and the unnormalized initial state is $\ketbra{\phi^{(\text{i})}}{\phi^{(\text{i})}}= \prod_{m} (1+i \gamma_{2m-1}\gamma_{2m})$.
While the final states $\final$ [Eq.~\eqref{eq:final_states}] are not Gaussian, we use 
\begin{align}
 | \calZ |^2 =  \bra{\phi^\text{(i)}} \calM^\dagger \underbrace{ \frac{\ket{\phi^\text{(f)}_q} \bra{\phi^\text{(f)}_q} + \ket{\tilde{\phi}^\text{(f)}_q} \bra{\tilde{\phi}^\text{(f)}_q}}{2} }_{\eqqcolon \rho^\text{(f)}_q} \calM \ket{\phi^\text{(i)}} ,
 \label{eq:absolute_value}
\end{align}
where $\ket{\tilde{\phi}^\text{(f)}_q}$ are the alternative final states, and
\begin{align}
 \rho^\text{(f)}_{q} = \frac{1}{2} \left[ 1+i(-1)^q \gamma_{1} \gamma_{2(M+1)} \mathcal{P} \right] \rho^\text{(f)}_\partial 
\end{align}
with the parity operator $\mathcal{P} = i^{M+1} \prod_{m=1}^{2(M+1)} \gamma_m$ and the unnormalized density matrix $\rho^\text{(f)}_\partial = \prod_{m=2}^{M} (1+i\gamma_{2m-1}\gamma_{2m})$ acting on Majoranas $3, \dots, 2M$.
We defined $\rho^\text{(f)}_\partial$ since the notion of a partial contraction on certain legs will be useful when discussing the interpretation as a quantum dot in Sec.~\ref{sec:quantum_dot}.
Although $\rho^\text{(f)}_q$ is %
not a Gaussian operator since it contains $\mathcal{P}$, we always contract it with the even-parity state $\calM \initial$, hence, we can simply replace $\mathcal{P} \mapsto 1$.

While the phase of $\calZ$ cannot be directly obtained from Eq.~\eqref{eq:absolute_value}, we can get the relative phase between $\mathcal{Z}_{0,s}$ and $\mathcal{Z}_{1,s}$ by choosing different final states 
\begin{align}
 \rho^\text{(f)}_+ &= (1+i\gamma_{2M+1}\gamma_{2M+2}) \rho^\text{(f)}_\partial  , \\
 \rho^\text{(f)}_y &= (1 - i \gamma_{1}\gamma_{2M+1} \mathcal{P}) \rho^\text{(f)}_\partial 
\end{align}
such that  %
\begin{align}
 \bra{\phi^\text{(i)}} \calM^\dagger \rho^\text{(f)}_{+} \calM \ket{\phi^\text{(i)}}
 &= \frac{p(s)}{2} %
 \left( 1 + \sin \theta_s \cos \varphi_s \right) \label{eq:GaussianX} \\
 \bra{\phi^\text{(i)}} \calM^\dagger \rho^\text{(f)}_{y} \calM \ket{\phi^\text{(i)}}
 &= \frac{p(s)}{2} %
 \left( 1 + \sin \theta_s \sin \varphi_s \right) \label{eq:GaussianY}
\end{align}%
with the parametrization from Eqs.~\eqref{eq:Z0param} and~\eqref{eq:Z1param}.
Using this method, we can compute $p(s)$, %
$\varphi_s \in [0,2\pi)$, and $\theta_s \in [0,\pi/2]$; recall that $\theta_s$ is restricted since we employ maximum-likelihood decoding, which implies $|\mathcal{Z}_{0,s}|\ge |\mathcal{Z}_{1,s}|$ and hence $\theta_s \le \pi/2$.

To sample Pauli error-strings $\eta$ and hence the syndromes $s$, we employ a variation of the algorithm by \citeauthor{Bravyi:2018ea}~\cite{Bravyi:2018ea} (also cf. Ref.~\onlinecite{Venn:2020ge}) that is suitable for the construction of the network model described here.
We describe the algorithm in App.~\ref{app:sampling} and provide more details on the Gaussian circuit in App.~\ref{app:flo}.

\subsection{Embedding the logical state}
\label{sec:embedLogical}

In Ref.~\onlinecite{Bravyi:2018ea}, \citeauthor{Bravyi:2018ea} showed that the logical information of the final states of a square-lattice SC, under similar error and recovery channels, is encoded in the state of 4 ``dangling'' Majorana modes of a Gaussian TN; see also Ref.~\onlinecite{Venn:2020ge}. Here, by bending the legs corresponding to these Majoranas [Fig.~\ref{fig:transfer_mapping}(c)], we explicitly construct a 4-Majorana state using the (1+1)D circuit above.

We hence show that the logical states are encoded into the (1+1)D circuits from the previous section. To this end, instead of contracting the evolved state $\calM \initial$ with one of the final states $\final$, we contract it only with the state $\ket{ \phi^\text{(f)}_\partial } = \sqrt{2}^{M-1} \bigotimes_{m=2}^{M} \ket{+}_m$. The outgoing legs (or Majorana lines) of qubits $m = 2, \dots, M$ are contracted with $\ket{+}_m$, whereas legs of qubits $1$ and $M+1$ are left uncontracted, leading to the 2-qubit state [see Fig.~\ref{fig:transfer_mapping}(c)]
\begin{align}
 \ket{\phi_{s}} = \left( \id_1 \otimes \bra{\phi^\text{(f)}_\partial} \otimes \id_{M+1} \right)  \calM \ket{ \phi^\text{(i)}} . \label{eq:partial_contr}
\end{align}

This (unnormalized) state $\ket{\phi_s}$ encodes the final logical state of the SC.
Due to parity conservation, $\ket{\phi_{s}}$ is a $+1$ eigenstate of $S = \tau^x \otimes \tau^x$; hence, $\ket{\phi_{s}}$ is a two-qubit phase-flip code state. 
Recalling that choosing $\ket{00}$ (or $\ket{11}$) and $\ket{01}$ (or $\ket{10}$) for qubits $1$ and $M+1$ leads to amplitudes ${\cal Z}_{0,s}$ and ${\cal Z}_{1,s}$, respectively, we identify $\tilde{Z} = \tau^z \otimes \tau^z$ as the logical operator with eigenstates $\ket{\tilde{0}} = (\ket{00} + \ket{11})/\sqrt{2}$ and $\ket{\tilde{1}} = (\ket{01} + \ket{10})/\sqrt{2}$. The other logicals can be chosen $\tilde{X} = \tau^x \otimes \id$ and $\tilde{Y} = \tau^y \otimes \tau^z$; there also exist equivalent logicals $\tilde{\sigma}S$ for $\tilde{\sigma}=\tilde{X}, \tilde{Y}, \tilde{Z}$. In terms of this two-qubit code, we thus have 
\begin{align}
    \ket{\phi_s} = {\cal Z}_{0,s} \ket{\tilde{0}} + {\cal Z}_{1,s} \ket{\tilde{1}}.
\end{align}This demonstrates the embedding of the SC logical state $\kpsis \propto {\cal Z}_{0,s} \zeroL + {\cal Z}_{1,s} \oneL$ [Eq.~\eqref{eq:psi_s}] into the quantum circuit's Hilbert space.

\begin{figure}[tp]
\centering
  \includegraphics[width=\columnwidth]{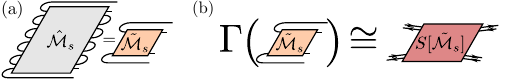}
  \caption{%
  (a): The full quantum circuit $\calM$ acts on $2M$ Majoranas (wires) and leaves the first and last modes untouched before being contracted with $\ket{\phi^{(f)}_\partial}$, giving the four-Majorana state $\ket{\phi_s}$. This Gaussian state $\ket{\phi_s}$ is equivalently obtained by acting with a Gaussian gate $\tilM$ on a 4-Majorana state. (b): The ensemble of correlation matrices $\Gamma(\ket{\phi_s})$ is isomorphic to that of the scattering matrices $S[\tilde{\cal M}_s]$ of a quantum dot, with incoming and outgoing Majorana modes indicated by arrows.}
\label{fig:quantum_dot}
\end{figure}

Since $\ket{\phi_s}$ and the initial state $\ket{++}$ on the ancilla qubits $1$ and $M+1$ are (unnormalized) Gaussian states, it is instructive to view the net effect of Eq.~\eqref{eq:partial_contr} as a Gaussian operation via [see Fig.~\ref{fig:quantum_dot}(a) for a depiction]
\begin{align}
    \frac{\ket{\phi_s}}{\sqrt{\braket{\phi_s|\phi_s}}} \eqqcolon \tilM \ket{++},
\end{align}with the 2-Majorana gate $\tilde{{\cal M}}_s(\gamma_1, \gamma_2)$ reading 
\begin{align}
    \tilde{{\cal M}}_s = \frac{ \cos\frac{\theta_s}{2} (1+i\gamma_1 \gamma_2) + \sin\frac{\theta_s}{2} e^{i\varphi_s}  (1-i\gamma_1 \gamma_2) }{\sqrt{2}} . \label{eq:tildeM}
\end{align}This effective gate $\tilde{{\cal M}}_s$ also encodes the final logical state of the SC, but it moreover enables us to construct a quantum dot scattering problem, as we next explain.

\section{Logical state via symmetric quantum dot}
\label{sec:quantum_dot}

In this Section, we explain how the projected logical ensemble corresponds to an ensemble of scattering matrices of a symmetric quantum dot. 
To construct this dot, in Sec.~\ref{sec:scatt_net}, we map the bulk of the quantum circuit from Fig.~\ref{fig:transfer_mapping} to a Majorana scattering network, which we identify with the internal fabric of the dot [Fig.~\ref{fig:overview}(b)]. In Sec.~\ref{sec:qdot_scatt}, by interpreting the uncontracted modes from Fig.~\ref{fig:quantum_dot} as the leads of the dot, we then find the dot's scattering matrix [see also Fig.~\ref{fig:overview}(b)]. In Sec.~\ref{sec:symm_lattices}, we discuss the symmetry classes of the network model, and thus the quantum dot, based on the surface code lattice. Finally, in Sec.~\ref{sec:implications}, we examine the implications of a chaotic quantum dot on the PLE.

\subsection{Fabric of quantum dot: Bulk scattering network}
\label{sec:scatt_net}

Here we largely follow Ref.~\cite{Jian:2022jg} to construct a charge-conserving network model.
We start by noting that the operators $\hat{V}^{(l,m)}$ [Sec.~\ref{sec:SC_to_circuit}] that comprise the Gaussian circuit $\calM$ transform single Majorana operators linearly
\begin{align}
\hat{V}^{(l,m)} \ \boldsymbol \gamma \ [\hat{V}^{(l,m)}]^{-1} = \boldsymbol\gamma^T \underbrace{ \begin{pmatrix}
 -\frac{i}{\tan (2 \phi )} & \frac{\eta^{(l,m)}}{\sin (2 \phi )} \\
 -\frac{\eta^{(l,m)}}{\sin (2 \phi )} & -\frac{i}{\tan (2 \phi )} \end{pmatrix}}_{t_v}
 \label{eq:tv_definition}
\end{align}
where $\boldsymbol\gamma = (\gamma_{2m},\gamma_{2m+1})^T$, and where $t_v$ can be written as an exponential of an anti-symmetric (but generally complex-valued) matrix; hence it satisfies $t_v^T = t_v^{-1}$.
We now introduce the time-reversed copy $t_v^*$  and the doubled transfer matrices~\cite{Jian:2022jg} (the doubling is necessary for lattices with odd coordination number~\cite{Jian:2022jg,fvjbbb_prl}; see also Sec.~\ref{sec:symm_lattices})
\begin{align}
\mathsf{t}' [v]= \begin{pmatrix} t_v & 0 \\ 0 & t_v^* \end{pmatrix},
\end{align}
which is pseudo-unitary since it satisfies
 $(\mathsf{t}'[v])^{-1}= \Sigma_x (\mathsf{t}'[v])^\dagger \Sigma_x ,$
where $\Sigma_x = \begin{psmallmatrix}
    0 & \id \\ \id & 0
\end{psmallmatrix}$ acts on the outer grading of time-reversed copies; hence, we can interpret $\mathsf{t}'[v]$ as a transfer matrix~\cite{Beenakker:1997gz}.
It helps to work in a rotated basis
\begin{align}
\mathsf{t}[v]
&=\frac{1}{2}\left(1 + i \Sigma_x \right) \mathsf{t}'[v] \left(1-i\Sigma_x \right) \\
&=\frac{1}{2}
\begin{pmatrix}
 t_v + t_v^* & -i \left(t_v-t_v^*\right) \\
 i \left(t_v-t_v^*\right) & t_v + t_v^*
\end{pmatrix},
\end{align}
which satisfies $\mathsf{t}^{-1} = \Sigma_z  \mathsf{t}^\dagger  \Sigma_z$.

In this basis, we can relate the $2\times 2$ blocks of the transfer matrix to transmission and reflection matrices of a scattering matrix $S [\hat{V}] = \begin{psmallmatrix}
    r & t' \\ t & r'
\end{psmallmatrix}$~\cite{Beenakker:1997gz}, i.e., identify the blocks~\cite{Jian:2022jg} 
\begin{align}
\mathsf{t} [v] \equiv
\begin{pmatrix}
 {t^\dagger }^{-1} & r' {t'}^{-1} \\
 -{t'}^{-1} r & {t'}^{-1}
\end{pmatrix}
= \begin{pmatrix}
 \frac{i \eta}{\sin (2 \phi )} \tau^y & \frac{-1}{\tan (2 \phi )} \id \\
 \frac{1}{\tan (2 \phi )} \id & \frac{i \eta}{\sin (2 \phi )} \tau^y
\end{pmatrix}
\end{align}
hence $t=i \eta \sin (2 \phi ) \tau^y$, $t'=-i \eta \sin (2 \phi ) \tau^y$ and $r=r'=i \eta \cos (2 \phi ) \tau^y$, with Pauli $\tau^\mu$ acting on the space spanned by Majoranas $(2m,2m+1)$. One can verify that $S$ is unitary $S^\dagger S = \id$ and real $S= S^*$, and thus is a Majorana scattering matrix.
One can also show that this scattering problem is in class DIII by checking~\cite{Jian:2022jg}: (\textit{i}) time-reversal symmetry $\Sigma_{\rm TR}^\dagger \mathsf{t}^* \Sigma_{\rm TR} = \mathsf{t}$ with $\Sigma_{\rm TR} = \begin{psmallmatrix}
    0 & - \id  \\ \id  & 0
\end{psmallmatrix}$ and $\Sigma^2_{\rm TR} = -\id$, (\textit{ii}) particle-hole symmetry $\Sigma_{\rm PH}^\dagger \mathsf{t}^* \Sigma_{\rm PH} = \mathsf{t}$ with $\Sigma_{\rm PH} = \begin{psmallmatrix}
    0 &  \id  \\ \id  & 0
\end{psmallmatrix}$ and $\Sigma^2_{\rm PH} = \id$, and (\textit{iii}) chiral symmetry $\Sigma_{\rm C}^\dagger \mathsf{t} \Sigma_{\rm C} = \mathsf{t}$ with~$\Sigma_{\rm C} = \begin{psmallmatrix}
    \id & 0  \\ 0 & -\id
\end{psmallmatrix}$ and $\Sigma^2_{\rm C} = \id$. (Class DIII can also be identified from $S$ being antisymmetric besides being real and unitary~\cite{bardarson2008proof,Fulga_PhysRevB.83.155429,Fulga_PhysRevB.86.054505}.)

The resulting bulk scattering network is shown in Fig.~\ref{fig:overview}(b); its scattering matrix $S[\calM]$ can be constructed from the single scattering center matrices $S[\hat{V}]$, cf.~\cite{fvjbbb_prl, Jian:2022jg}. We next identify this bulk network as the internal fabric of a synthetic quantum dot. %

\subsection{Scattering matrix of quantum dot} \label{sec:qdot_scatt}

We now connect the effective gate $\tilde{{\cal M}}_s$ [Eq.~\eqref{eq:tildeM}, Fig.~\ref{fig:quantum_dot}] with the scattering matrix of a quantum dot. We proceed analogously to Sec.~\ref{sec:scatt_net}, focusing on $\tilde{\cal M}_s$ instead of $\hat{V}^{(l,m)}$; however, here, as we discuss below, the scattering modes are associated with the leads of a quantum dot instead of modes of the bulk network [cf. Fig.~\ref{fig:overview}(b)]. By considering the single-particle action of $\tilde{\cal M}_s$, we find (as detailed in App.~\ref{app:SfromM})%
\begin{align}
 S[\tilde{\cal M}_s] = \begin{pmatrix}
        r_s & t'_s \\ t_s & r'_s
    \end{pmatrix}, \label{eq:sdot}
\end{align}
with $r_s = r'_s = \cos \theta_s e^{i (\pi/2) \tau^y}$ and $t_s = t'^T_s = \sin \theta_s e^{i \varphi_s \tau^y}$ for Pauli $\tau^y$. As done in Sec.~\ref{sec:scatt_net}, we can show that $S[\tilde{\cal M}_s]$ is a real scattering matrix in class DIII. Moreover, $S[\tilM]$ is isomorphic to the correlation matrix of the final state $\Gamma(\ket{\phi_s})$ through a fixed (i.e., $s$-independent) isomorphism [as detailed in App.~\ref{app:isomorph} and shown in Fig.~\ref{fig:quantum_dot}(b)]
\begin{align}
    S[\tilM] \cong \Gamma(\ket{\phi_s}). \label{eq:SisG}
\end{align}

This a key result of our work: The projected logical ensemble of final states of the SC %
$\ens = \{ p(s|\rho), %
\kpsis\}$ [Eq.~\eqref{eq:LE}, with $\rho = \ketbra{\overline{0}}{\overline{0}}$ such that $p(s|\rho) = p(s)$] is in a one-to-one correspondence to this ensemble of scattering matrices
$\{ p(s), %
S[\tilde{\cal M}_s] \}$ [Eq.~\eqref{eq:sdot}].
The Altland-Zirnbauer symmetry class of $S[\tilde{\cal M}_s]$ is determined by the SC and coherent rotation model; here, for the honeycomb SC with single-qubit $X$ rotations, this is class DIII, as explicitly shown above.

Thus far we have interpreted $S[\tilM]$ as a $4 \times 4$ class DIII scattering matrix. We next show that $S[\tilM]$ is the scattering matrix of a quantum dot with the scattering network as its fabric.
To this end, we relate transport in the network model to the overall transport through the dot. %
We consider three scattering regions as shown in Fig.~\ref{fig:overview}(b):
The left and right leads 
are described by the scattering matrices $S_{\rm L}$ and $S_{\rm R}$, respectively, with
\begin{align}
 S_{\rm L} = \begin{pmatrix}  \mathds{O}_2 & \mathcal{T}^T \\ \mathcal{T} & \mathcal{R} \end{pmatrix} , & & 
 S_{ \rm R} = \begin{pmatrix} \mathcal{R} & \mathcal{T} \\ \mathcal{T}^T &  \mathds{O}_2 \end{pmatrix},
\end{align}
where $\mathds{O}_d$ is a $d \times d$ matrix filled with zeros, with the $2M \times 2M$ matrix $\mathcal{R} = \mathds{O}_1 \left[ \oplus_{i=1}^{M-1} \begin{psmallmatrix} 0 & 1 \\ -1 & 0 \end{psmallmatrix} \right]  \oplus \mathds{O}_1$,
and the $2M \times 2$ matrix $\mathcal{T}$ whose entries are $\mathcal{T}_{ij} = \delta_{i,1}\delta_{j,1} + \delta_{i,2M}\delta_{j,2}$; ${\cal R}$ and $\cal T$ follow from the Majorana representation of the single-qubit initial $\initial_m$ and final $\ket{\phi^{(\rm f)}_\partial}_m$ states.
The left leads are perfectly transmissive for the two incoming modes from the left, and reflect all other modes coming from the right. The right leads reflect all but two modes from the left, and are perfectly transmissive for the other two modes.
Concatenating~\cite{Tamura:1991ki} the boundary scattering matrices with the bulk scattering matrix yields
\begin{align}
 S [\tilM] = S_{\rm R} \circ S [\calM] \circ S_{ \rm L} ,
\end{align}
recovering the same $S[\tilM]$ as in Eq.~\eqref{eq:sdot}.

\subsection{Symmetry classes and surface code lattices}
\label{sec:symm_lattices}

\begin{figure}[tp]
\centering
  \includegraphics[scale=1]{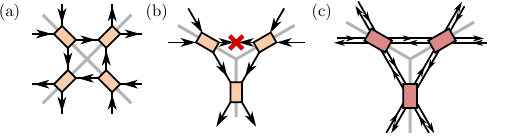}
  \caption{(a): The gates $\hat{V}^{(l,m)}$ act unitarily along the edges of the $S_v^X$ stabilizer lattice, which is indicated by the arrows that denote one of two possible choices for the direction of a unitary evolution.
  For lattices with only even-weight $S_v^X$, the lattice of $\hat{V}^{(l,m)}$ gates can be directly interpreted as a charge-conserving network model in symmetry class D, as shown in Ref.~\onlinecite{fvjbbb_prl}.
  (b): For lattices with odd-weight $S_v^X$, such a construction leads to a contradiction, as indicated by the red cross: incoming and outgoing modes cannot be connected. Following Ref.~\onlinecite{Jian:2022jg}, we add a time-reversed copy to obtain a charge-conserving network in symmetry class DIII (see also Ref.~\cite{Yang:2026ktu}).
  }
\label{fig:lattices_networks}
\end{figure}

For lattices with only even-weight $S_v^X$ stabilizers, i.e., surface codes defined on lattices with even coordination number, the quantum circuit $\calM$ can be readily interpreted as a class D network model for transport~\cite{fvjbbb_prl}.
Crucial for this identification is that the single-particle matrix $t_v$ defined in Eq.~\eqref{eq:tv_definition} satisfies $\tau^x t_v^\dagger \tau^x = - t_v^{-1}$, i.e., $t_v$ is pseudo-unitary~\cite{Beenakker:1997gz} up to a minus sign~\cite{fvjbbb_prl}.
This is equivalent to being unitary up to permuting indices, such that the top modes become the incoming modes and the bottom modes the outgoing modes of a corresponding unitary matrix; incoming and outgoing modes can also be exchanged in this picture.
(Such a reshuffling of indices is commonly used in dual-unitary circuits~\cite{Bertini2019} where operators that are unitary before and after the reshuffling are considered.)
We visualize this in Fig.~\ref{fig:lattices_networks}(a): the arrows indicate the direction of the unitary evolution, which is always along the lattice of $S_v^X$ stabilizers.
As indicated in Fig.~\ref{fig:lattices_networks}(a), for lattices with even coordination number, the arrows in a whole network can be consistently matched, such that a direct construction of a charge-conserving network model is possible without the doubling described in Sec.~\ref{sec:scatt_net}.
If one were to do the doubling anyway, one would find that the scattering matrix splits into two uncoupled blocks, which follows from the fixed phases $\varphi_s$; in App.~\ref{app:dot_scatt_mat}, we give such an example.

For lattices with odd coordination number, however, doubling the degrees of freedom is necessary for a successful identification with a current-conserving network model~\cite{Jian:2022jg}.
We show an example for a vertex of such a lattice in Fig.~\ref{fig:lattices_networks}(b): again viewing $t_v$ as a unitary matrix along the direction of the $S_v^X$ stabilizer lattice, we find that the arrows cannot be matched up consistently, always leading to a contradiction as indicated by the cross in Fig.~\ref{fig:lattices_networks}.
(One can understand this in terms of frustration: each edge carries arrows pointing towards or away from the node. For charge-conserving models, for each node the number of arrows pointing towards and away from it must be equal, which is not possible for odd coordination numbers.)
Doubling the degrees of freedom, as in Ref.~\cite{Jian:2022jg}, is thus necessary for a charge-conserving model. We show an example for a degree-three vertex in Fig.~\ref{fig:lattices_networks}(c).
The difference between the symmetry classes for lattices with only even-weight $S_v^X$ stabilizers and odd-weight stabilizers has also been noted in recent works~\cite{Yang:2026ktu,yan2026nonlinearsigmamodelsurface}.

The symmetry class determines the large-scale transport properties of the corresponding scattering network.
In the metallic regime, a NL$\sigma$M~\cite{Wegner1979,Efetov1980,ReadLudwig2000, AndersonTransRMP} is the effective long-wavelength field theory for transport~\cite{AndersonTransRMP}.
As effective theories, NL$\sigma$Ms depend only on the symmetric space in which they are defined, which in turn reflects the symmetry class of an underlying Hamiltonian or network model~\cite{AndersonTransRMP}.
Originally derived using the replica trick~\cite{Wegner1979}, the limit where the number of replicas $N$ goes to zero is taken when the disorder is distributed independently from the system (e.g., Gaussian disorder).

For the network model describing coherent errors, a subtlety arises~\cite{yan2026nonlinearsigmamodelsurface, Yang:2026ktu}:
disorder realizations $\{\eta\}$ correspond to syndromes sampled according to their measurement probabilities, which in turn can be understood as a Born-rule sampling of nonunitary gates in the corresponding quantum circuit; cf.\ App.~\ref{app:sampling}.
As shown in 
Refs.~\onlinecite{NLSM1,NLSM2, NLSM3}, such Born-rule sampling necessitates taking a different replica limit, namely $N \to 1$.
For coherent errors, sampling Pauli strings (and hence $\eta$) according to their probabilty thus corresponds to the $N\to 1$ replica limit~\cite{yan2026nonlinearsigmamodelsurface, Yang:2026ktu}, whereas sampling $\eta$ uniformly~\cite{jbfvbb_prr} (or according to a suboptimal decoder~\cite{yan2026nonlinearsigmamodelsurface}) is described by the $N\to 0$ limit.
In Sec.~\ref{sec:HSCthreshold} and App.~\ref{app:TwirlNumerics}, we shall compare the predictions of these NL$\sigma$Ms with the numerically observed transport properties of the scattering network.

\subsection{Chaotic quantum dot implications on the projected logical ensemble}
\label{sec:implications}

Chaotic quantum dots can exhibit universal behavior, wherein their scattering matrices are distributed according to symmetry-dependent ensembles from RMT~\cite{Beenakker:1997gz,AlhassidRMP}. For disordered quantum dots, this regime requires the dot to be a diffusive metallic grain, i.e., its linear size $L$ to exceed the mean free path, and $L\ll \xi_{\rm loc}$ to hold, where  $\xi_{\rm loc}$ is the localization length. For quantum transport to be chaotic, $t_{\rm dw}/t_{\rm erg}\gg 1$ should furthermore hold for the ratio of the dwell time $t_{\rm dw}$ over which a particle escapes the dot and the ergodic time $t_{\rm erg}$ needed for a particle to explore the dot. For few-mode contacts, as in our settings, $t_{\rm dw}/t_{\rm erg}\sim g(L)$, where $g(L)$ is the finite-size scaling of the bulk conductivity~\cite{edwards1972numerical,Thouless77_PhysRevLett.39.1167,Abrahams79_PhysRevLett.42.673,akkermans2007mesoscopic,Beenakker:1997gz,AlhassidRMP}. The corrections to the RMT behavior decay polynomially in $g(L)$, as was shown for various quantities using supersymmetric NL$\sigma$M techniques~\cite{altshuler1986repulsion,efetov1983supersymmetry,Fyodorov_95_PhysRevB.51.13403,efetov1999supersymmetry,mirlin2000statistics,bocquet2000disordered}.

In our setting, the relevant dots belong to classes D and DIII, corresponding respectively to lattices with only even-weight and those that include odd-weight $S^X_v$ stabilizers. Above the QEC threshold, the associated network model is expected to enter its delocalized phase for class DIII and was shown to have $\xi_{\rm loc} \gg L$ for class D (cf. Ref.~\onlinecite{fvjbbb_prl} for the square-lattice SC and App.~\ref{app:classD} for the triangular-lattice SC). In the diffusive regime, we thus expect convergence to a chaotic quantum dot with emergent RMT universality. This is a phenomenological expectation that is natural based on a diffusive metallic grain. (Mathematically, this amounts to the assumption that the diffusive metal phenomenology allows the results from the previous paragraph, which describe the $N\to 0$ replica limit, to carry over to $N\to 1$.) 

This emergent universality enables us, via the dot to logical states correspondence $S[\tilM] \leftrightarrow \kpsis$, to predict an emergent universality for the PLE. Since $S[\tilM]$ and $\Gamma_s$ are isomorphic through a fixed ($s$-independent) isomorphism, RMT universality implies that $S[\tilM]$ and $\Gamma( \ket{\phi_s})$ are uniformly distributed in isomorphic symmetric spaces~\cite{AltlandZirnbauer}. Focusing on $\Gamma( \ket{\phi_s})$, the symmetric spaces in class D and DIII are ${\rm O}(2)\setminus {\rm SO}(2)$
and $\frac{{\rm SO}(4)}{{\rm U}(2)}$, respectively~\cite{AltlandZirnbauer}; hence, the probability density of the logical angles is $p(\theta_s, \varphi_s) \propto [\delta(\varphi_s - \varphi_0) + \delta(\varphi_s - \varphi_0 - \pi)]\Theta(\pi/2 - \theta_s)$ for class D ($\varphi_0 \in \{0, \pi/2 \}$, cf. Sec.~\ref{sec:LE}), and $p(\theta_s, \varphi_s) \propto \sin \theta_s \Theta(\pi/2 - \theta_s)$ for class DIII, with Heaviside function $\Theta$. These imply that $\kpsis$ is uniformly distributed over a Bloch (semi)circle and hemisphere for class D and DIII, respectively. As discussed in the next section (class DIII) and App.~\ref{app:classD} (class D), we numerically find results consistent with this RMT and PLE universality being approached for SC in a regime above the QEC threshold.

\section{Symmetry class DIII numerics} \label{sec:BornNumerics}

In this Section, we discuss numerical results for a surface code in symmetry class DIII, while we present results for class D in App.~\ref{app:classD}. For the simulations, we focus on the honeycomb SC with length $L$, width $M=L$, rotation angle $\phi$, and initial state $\kpsi = \ket{\overline{0}}$ [Sec.~\ref{sec:SCsetup}, Fig.~\ref{fig:overview}(a)]. We use free-fermion-based algorithms to efficiently contract the various (1+1)D Gaussian circuits [Sec.~\ref{sec:GaussianCirc}, App.~\ref{app:flo}]. Using this, we first sample syndromes $s$ according to their probability %
$p(s|\rho)$, and then compute the overlaps via Eqs.~\eqref{eq:absolute_value},  \eqref{eq:GaussianX}, and \eqref{eq:GaussianY}, %
\begin{align}
    p_{0,s} = \left|\braket{\overline{0}|C_s U |\overline{0}}\right|^2 &= |{\cal Z}_{0,s}|^2 = \frac{p(s)}{2}(1 + \cos \theta_s) , \label{eq:pz0} \\
    p_{1,s} = \left| \braket{\overline{1}|C_s U |\overline{0}}\right|^2 &= \left| {\cal Z}_{1,s}\right|^2 = \frac{p(s)}{2}(1 - \cos \theta_s), \\
    \left| \braket{+\XL|C_s U |\overline{0}}\right|^2 &=  \frac{p(s)}{2}(1 + \sin \theta_s \cos \varphi_s), \\
    \left| \braket{+\YL|C_s U |\overline{0}}\right|^2 &=  \frac{p(s)}{2}(1 + \sin \theta_s \sin \varphi_s), \label{eq:py}
\end{align}where $\ket{+\XL} = (\ket{\overline{0}} + \ket{\overline{1}})/\sqrt{2}$ and $\ket{+\YL} = (\ket{\overline{0}} + i\ket{\overline{1}})/\sqrt{2}$.  Using the MLD, we reconstruct the logical state $\kpsis = \cos(\theta_s/2) \ket{\overline{0}} + e^{i\varphi_s}\sin(\theta_s/2) \ket{\overline{1}}$ [cf. Eq.~\eqref{eq:psi_s}], 
with angles $\theta_s \in [0, \frac{\pi}{2}]$ and $\varphi_s \in [0,2\pi).$ We also compute the bulk conductivity of the network model. Altogether, we can estimate the logical error rate, the bulk conductivity, the quantum dot conductance (including its Wasserstein-1 distance), and the distances between finite moments of the logical and Haar ensembles, as we discuss next.

\subsection{QEC threshold and insulator-to-metal transition} \label{sec:HSCthreshold}

\begin{figure}[tp]
    \centering
    \includegraphics[width=\linewidth]{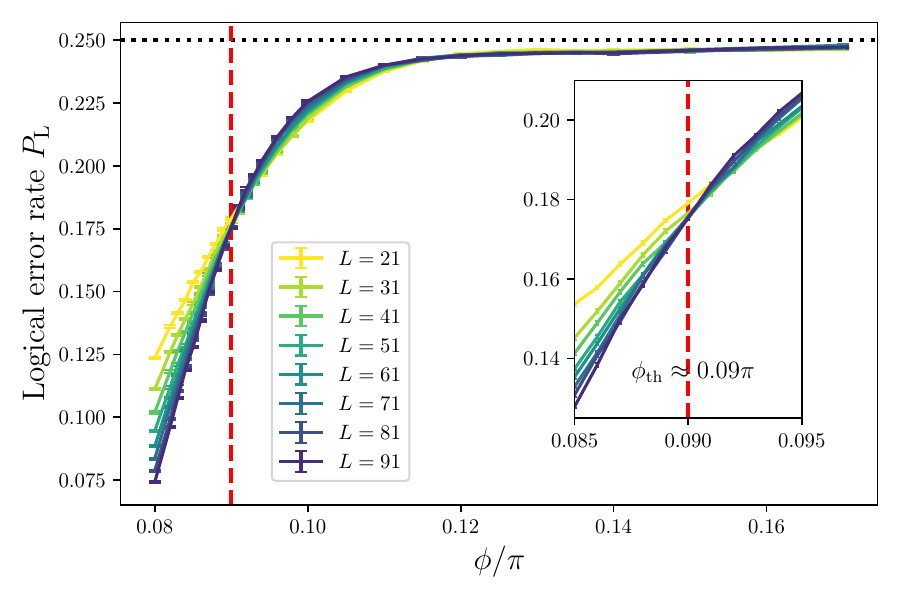}
    \caption{Logical error rate $P_{\rm L}$ vs coherent error angle $\phi$ for the honeycomb surface code with length $L$. $P_{\rm L}$ decays with $L$ below the error threshold $\phi_{\rm th}$ (vertical dashed red line). The inset shows data closer to the QEC transition. For $\phi > \phi_{\rm th}$, $P_{\rm L}$ initially increases with $L$, after which it becomes weakly dependent on $L$ for the accessible system sizes. The horizontal dotted black line shows the minimum average infidelity for post-QEC states Haar-distributed over a Bloch hemisphere. Error bars show the standard error of the mean (SE). Data averaged over $10^5$ to $10^6$ syndromes.}
    \label{fig:err_rate}
\end{figure}

To estimate the QEC threshold of this code, we consider the logical error rate %
\begin{align}
    P_{\rm L} \coloneqq \sum_s p(s) \min_q \frac{p_{q,s}}{p(s)} = \sum_s \min_q p_{q,s},%
    \label{eq:PL}
\end{align}which quantifies how distinct the two correction classes are on average, with probabilities $p_{q,s} = \vert \calZ \vert^2$, and using the maximum-likelihood decoder  [Sec.~\ref{sec:SCsetup}], reduces to $P_{\rm L} = \sum_s p(s) \sin^2(\theta_s/2)$. For coherent errors, $P_{\rm L}$ sets the minimal average infidelity between pre- $\ket{\psi_0}$ and post-QEC $\kpsis$ states~\cite{Venn:2020ge}.
For $\phi$ below the error threshold $\phi < \phi_{\rm th}$, $p_{0,s} \gg p_{1,s}$ for sufficiently large code distances $d$, and applying the correction $C_s$ yields $\kpsis$ close to $\ket{\psi_0}$ with high probability; an error-correcting phase can be defined as a regime where $P_{\rm L}$ decays exponentially with the code distance $d$. By contrast, the post-QEC $\kpsis$ is often far from $\ket{\psi_0}$ for $\phi > \phi_{\rm th}$; one thus generally does not expect $P_{\rm L}$ to decay with the code distance in such~phases.

We numerically find, using the logical error rate, that $\phi_{\rm th} = (0.09 \pm 0.001)\pi$ %
for the honeycomb SC [Fig.~\ref{fig:err_rate}]. We observe that, for $\phi < \phi_{\rm th}$, $P_{\rm L}$ decreases exponentially with the code distance $d\propto L$ (not shown), %
whereas for $\phi > \phi_{\rm th}$, data is consistent with $P_{\rm L}$ increasing with $L$ for a while %
after which it appears to become weakly-dependent on $L$, at least for the largest numerically accessible $L$. This weakly-$L$-dependent value approaches $P_{\rm L} \simeq 0.25$, corresponding to post-QEC states uniformly distributed over the Bloch hemisphere (and initial state $\zeroL$)~\cite{Venn:2020ge}. %
The data thus suggests that, for $\phi>\phi_{\rm th}$, the PLE might approach the Haar ensemble over a Bloch hemisphere. 

While indicative of a QEC threshold, the weak dependence of $P_{\rm L}$ on the code distance for large $\phi$ suggests that another method to pinpoint the transition could be useful. We next discuss a method based on the network model's bulk conductivity and argue that the QEC transition corresponds to an insulator-to-metal transition.

\begin{figure}[tp]
    \centering
    \includegraphics[width=\linewidth]{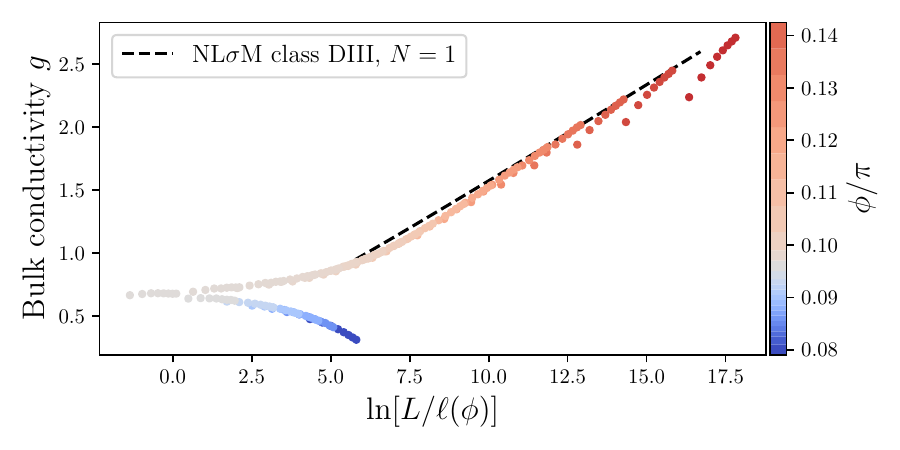}
    \caption{Bulk conductivity $g$ for the network model of the honeycomb surface code under coherent rotations. The system length and width are $21\leq L = M \leq 91$ with data averaged over $10^5$ to $10^6$ syndrome realizations. Error bars (SE) are imperceptible. $g$ scales with a $\phi$-dependent length $\ell(\phi)$. For the insulator $\phi < \phi_{\rm th}$, $g(L/\ell(\phi))$ decreases with $L$, with $\phi_{\rm th} = (0.089 \pm 0.002)\pi$. For the metal $\phi > \phi_{\rm th}$, $g$ increases and, for large $L$, is consistent with the scaling $g \propto (2\pi)^{-1}\ln[L/\ell(\phi)]$ of a nonlinear sigma model in class DIII in the replica limit $N\to1$ (dashed black line), see also~\cite{yan2026nonlinearsigmamodelsurface, Yang:2026ktu}.}
    \label{fig:bulk_g}
\end{figure}

\begin{figure*}[htp]
    \centering
    \includegraphics[width=\textwidth]{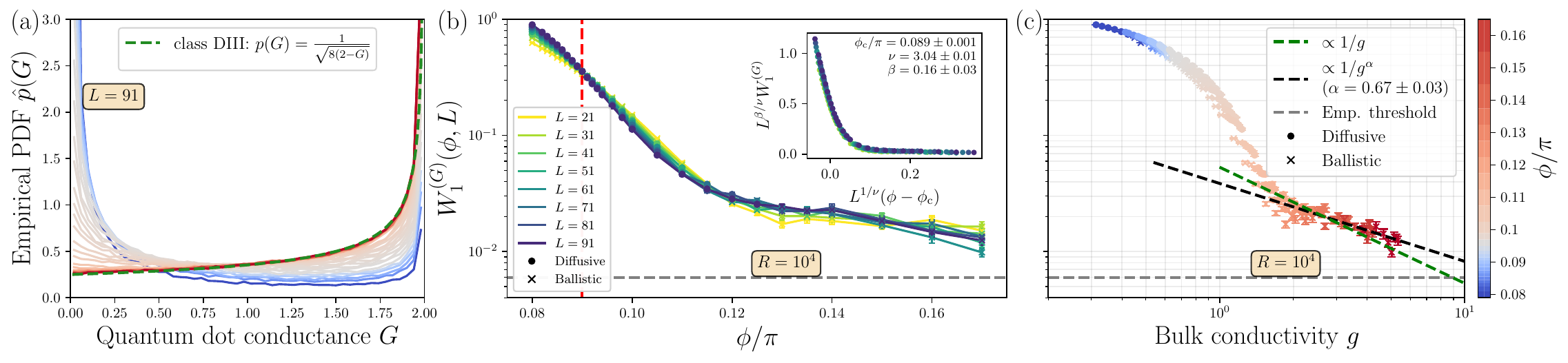}
    \caption{
    Emergent chaotic quantum dot and projected logical ensemble of the honeycomb surface code under coherent rotations. Data averaged over $10^5$ to $10^6$ syndrome realizations for length and width $21\leq L = M \leq 91$. Error bars show SE.
    (a): Empirical probability density function $\hat{p}(G)$ of the quantum dot's conductance $G$ for $L=91$. Above the error threshold $\phi > \phi_{\rm th} \approx 0.089\pi$ [vertical dashed red line in (b)], the empirical PDF approaches the random matrix theory prediction given by a dot made of a Majorana metal in symmetry class DIII (dashed green line). 
    (b): Wasserstein-1 distance $W^{(G)}_1(\phi,L)$ vs $\phi$ between quantum dot's and RMT distributions, or equivalently, the distributions of $G=2\sin^2\theta$ in the projected logical and Haar-hemisphere ensembles. Parameters $(L,\phi)$ that allow for a diffusive (ballistic) metal are shown as circles (crosses). Above threshold, in the diffusive regime, $W^{(G)}_1(\phi,L)$ decreases systematically with $L$ towards a value set by the number $R=10^4$ of syndromes per empirical CDF  [horizontal dashed gray line in (b) and (c)]. Inset: Finite-size scaling collapse suggestive of a continuous phase transition detecting a critical angle $\phi_{\rm c}$ consistent with the threshold $\phi_{\rm th}$. 
    (c): $W^{(G)}_1(\phi,L)$ vs bulk conductivity $g$ of the network model. In the metallic phase, $W^{(G)}_1(\phi,L)$ decays algebraically with $g$; the dashed green (black) line shows a linear (algebraic) fit.} %
    \label{fig:qdot}
\end{figure*}

We numerically find, using the bulk conductivity $g$ of the network model, that the honeycomb SC has an insulator-to-metal transition at $\phi_{\rm th}= (0.089 \pm 0.002)\pi$ [Fig.~\ref{fig:bulk_g}]. To quantify the transport through the bulk network, we write its scattering matrix in terms of $ M \times M$ reflection and transmission matrices (see also Sec.~\ref{sec:scatt_net})
\begin{align}
    S[\calM] = \begin{pmatrix}
        {\cal R}_s & {\cal T}'_s \\ {\cal T}_s & {\cal R}'_s
    \end{pmatrix}.
\end{align}The bulk conductivity is thus $g = (L/2M) \braket{\Tr [ {\cal T}^\dagger_s {\cal T}_s ]}$, where the average is over syndrome (i.e., disorder) realizations.
For sufficiently large $L$, we observe one-parameter scaling~\cite{LeeDisorderedSystems}, i.e., that the dimensionless $g$ scales with the ratio $L/\ell(\phi)$ of the length $L$ and a $\phi$-dependent length scale $\ell(\phi)$~\cite{fvjbbb_prl}:
for $\phi < \phi_{\rm th}$, $g(L/\ell(\phi))$ decreases exponentially with the code distance $d \propto L$, thus indicating that the network model localizes~\cite{PhysRevB.65.012506}; %
for $\phi > \phi_{\rm th}$, $g(L/\ell(\phi))$ increases and approaches the prediction $g \propto (2\pi)^{-1}\ln[L/\ell(\phi)]$ (cf. Ref.~\cite{AndersonTransRMP}) given by a diffusive metal described by a NL$\sigma$M in symmetry class DIII in replica limit $N\to1$ [Fig.~\ref{fig:bulk_g}], as expected from Sec.~\ref{sec:symm_lattices}. 
For $\phi \gtrsim 0.12\pi$ in the metallic phase, since the mean free path can increase significantly, it is reasonable to expect a crossover from a diffusive to a ballistic metal for finite-$L$ systems~\cite{PhysRevB.61.9690, PhysRevB.81.214203, Beenakker_2013, AndersonTransRMP}; we indeed find that only the largest $L$ ($L\geq 71$) data agrees well with the NL$\sigma$M prediction in this regime [cf. Fig.~\ref{fig:bulk_g}].
This transition yields an estimate $\phi_{\rm th}= (0.089 \pm 0.002)\pi$ of the QEC threshold consistent with the one based on the logical error rate [$(0.09 \pm 0.001) \pi$]; however, we note that the decrease and increase of $g$ with $L$ at a fixed $\phi$ provides a clearer way to identify the threshold.

\subsection{Emergent chaotic quantum dot} \label{sec:qdotNum}

Having established results suggesting that the network model transitions into a metallic phase above the QEC threshold, we now consider the effects of this bulk metal on the associated quantum dot [cf. Sec.~\ref{sec:quantum_dot}]. We focus on the dot's syndrome-dependent conductance [cf. Eq.~\eqref{eq:sdot}]
\begin{align}
    G_s = \Tr [ t^\dagger_s t_s] = 2\sin^2(\theta_s) \in [0,2].
\end{align}
We shall compare the statistics of $G_s$ with the RMT prediction for a chaotic quantum dot in symmetry class DIII. For two transmission channels, as our setting can be viewed, RMT respectively predicts the probability and cumulative density function (PDF and CDF) $\tilde{p}(G) = 1/\sqrt{8(2-G)}$ and $\tilde{P}(G) = 1 - \sqrt{(2-G)/2}$, cf. Ref.~\cite{dahlhaus_prb_2010}. 

We find that, in the bulk metallic phase, the syndrome-averaged data for the quantum dot agrees remarkably well with this RMT prediction [Fig.~\ref{fig:qdot}].
By sampling $R_{\rm tot}$ syndromes $\{s^{(i)}\}_{i=1}^{R_{\rm tot}}$, we construct an empirical distribution $\hat{p}(G)$; Fig.~\ref{fig:qdot}(a) shows such normalized histograms. Our data is consistent with $\hat{p}(G)$ approaching the PDF $\tilde{p}(G)$ from RMT above the error threshold $\phi > \phi_{\rm th}$.%

To probe this approach more carefully, we consider the Wasserstein-1 distance (which we review in App.~\ref{app:LErandomness}) $W^{(G)}_1$ between the quantum dot's and the RMT conductance probability distributions [Fig.~\ref{fig:qdot}(b,c), where we approximately identify parameters that allow for a diffusive or ballistic metal by $L> \ell(\phi)$ and $L<\ell(\phi)$, respectively\footnote{It might be appealing to increase the angle $\phi$ to access data with larger bulk conductivity $g(L,\phi)$. However, to avoid a crossover from diffusive to a ballistic metal~\cite{PhysRevB.61.9690, PhysRevB.81.214203, Beenakker_2013, AndersonTransRMP}, it would also be necessary to increase the system size $L$ since the mean free path is expected to increase with $\phi$ [similar to, but distinct from,~$\ell(\phi)$].}].
To estimate $W_1^{(G)}$, we first construct an empirical CDF $\hat{P}(G)$ from $R$ sampled syndromes $s^{(1)}, \dots, s^{(R)}$. Using this $\hat{P}(G)$ and the CDF $\tilde{P}(G)$ from RMT, we compute $W_1^{(G)}(\hat{P}, \tilde{P})$ via Eq.~\eqref{eq:W11D}, which we then average over $\approx R_{\rm tot}/R$ empirical CDFs, each constructed from $R$ syndromes, to estimate $W^{(G)}_1(P, \tilde{P})$. In the metallic phase $\phi > \phi_{\rm th}$, our data is consistent with $W^{(G)}_1(\phi, L)$ decaying algebraically with the bulk conductivity $g$ [Fig.~\ref{fig:qdot}(c)] towards a statistical threshold $\propto 1/\sqrt{R}$ for an empirical CDF built from $R$ syndromes~\cite{Fournier2015}. (In App.~\ref{app:addWDIII}, we process the same data but with a smaller $R$; there, the approach to this empirical threshold is more transparent.)
While we expected this in the diffusive regime, it appears to be a broader trend. 
Together with the other results in this section, this 
is suggestive of an emergent chaotic quantum dot above the code's error threshold, bringing us back to the question of how random might the PLE~be.%

\subsection{Emergent logical ensemble} \label{sec:emergentLHE}

\subsubsection{Wasserstein-1 distance} \label{sec:W1Born}

Here, we compare the logical ensemble $\ens$ [Eq.~\eqref{eq:LE}] with the Haar ensemble over the upper Bloch hemisphere $\ensH^*$, which is the appropriate ``maximally entropic'' ensemble within the constraints given by symmetry class DIII, maximum likelihood decoder, and initial state $\zeroL$ [Sec.~\ref{sec:implications}]. For convenience, we compare, using the Wasserstein-1 distance, the probability distributions of the variable $\vartheta = 2\sin^2\theta_s$ in $\ens$ and $\ensH^*$, which is equal to the conductance $G_s$ of the quantum dot. Since states $\kpsi$ distributed according to $\kpsi \sim \ensH^*$ have $p_{\rm H}(G(\kpsi)) = \tilde{p}(G)$, we have $W_1^{(G)}(p,\tilde{p}) =W_1^{(G)}(p,p_{\rm H})$, and we can thus simply reinterpret the data from the previous subsection.

We observe that $W^{(G)}_1(\phi,L)$ increases with $L$ below a critical angle $\phi < \phi_{\rm c}$, consistent with the expectation that below threshold the PLE concentrates at the initial logical state.
For $\phi > \phi_{\rm c}$, in the diffusive regime [shown with dots in Fig.~\ref{fig:qdot}(b)], $W^{(G)}_1(\phi,L)$ decays systematically with $L$ as expected from Sec.~\ref{sec:implications}. This decay is relatively slow, which is not surprising because the corrections to RMT are expected to scale as $\propto1/{\rm poly} [g(L,\phi)] \sim 1/{\rm poly}(\log L)$ for fixed $\phi$, cf. Sec.~\ref{sec:implications} and Sec.~\ref{sec:qdotNum} (and App.~\ref{app:TwirlNumerics} for numerics on systems described by the $N \to 0$ replica limit). For $\phi \gtrsim 0.12\pi$, where a crossover from a diffusive to ballistic metal is plausible (going beyond our current analytical expectations), $W^{(G)}_1(\phi,L)$ broadly decays with $\phi$ again towards the empirical threshold set by the number $R$ of syndromes per CDF. 
By performing a finite-size scaling analysis on the Wasserstein-1 distance $W^{(G)}_1(\phi, L)$~\cite{cardyFSS,andreas_sorge_2015_35293}, we find a critical angle $\phi_{\rm c} = (0.089 \pm 0.001)\pi$ consistent with the QEC threshold $\phi_{\rm th} \approx 0.089 \pi$ [Figs.~\ref{fig:err_rate} and~\ref{fig:bulk_g}]. This suggests that, when the quantum dot is chaotic (i.e., above threshold $\phi > \phi_{\rm th}$), the PLE $\ens$ approaches the class-DIII-constrained Haar ensemble~$\ensH^*$.

While the distribution of $G$ is a metric that can indicate how the PLE behaves, one would ideally like to consider the distribution of the states themselves. Although one may, in principle, numerically estimate the $W_1$ distance between the logical and Haar-hemisphere ensembles, evaluating such a $W_1$ is a formidable task. To this end, we instead consider finite moments of an PLE, thus assessing whether this forms state $k$-designs.

\subsubsection{Logical state designs}
\begin{figure}[tp]
    \centering
    \includegraphics[width=\linewidth]{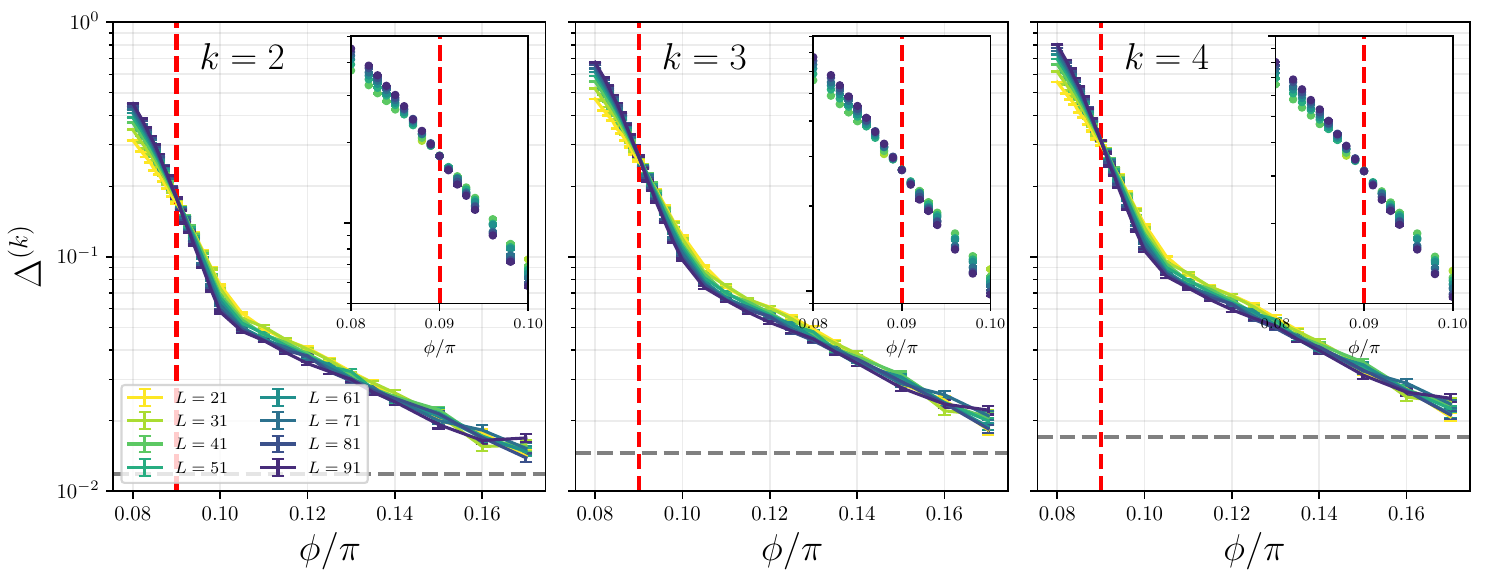}
    \caption{Projected logical ensemble $\cal E'$, assessed via trace distances $\Delta^{(k)}$  
     between the $k$th moments of $\cal E'$ and the Haar ensemble.
    We show data for the honeycomb surface code with rotation angle $\phi$, length $L$ and width $M=L$.
    Data averaged over $10^5$ to $10^6$ syndrome realizations. Error bars show SE.
     Above threshold (vertical dashed red line),  $\Delta^{(k)}$ appears to approach with $\phi$ a statistical threshold set by the number $R=10^4$ of syndromes per empirical $\cal E'$ (horizontal dashed gray line). For $(L,\phi)$ that allow for a diffusive metal (see inset), the approach is also systematic with $L$.
    }
    \label{fig:BornDesigns}
\end{figure}

Here, we consider a closely related logical ensemble~${\cal E'}$. To construct ${\cal E}'$, after applying the correction operation $C_s$ according to the maximum likelihood decoder [cf. Sec.~\ref{sec:SCsetup}], we apply with probability $1/2$ a logical $\XL$. Hence, the states in $\cal E '$ can be anywhere on the Bloch sphere for symmetry class DIII, unlike those in $\cal E$, which are constrained due to the MLD. The appropriate ``maximally entropic'' ensemble to compare with is thus the Haar ensemble over the full Bloch sphere $\ensH$. 

An ensemble of pure states whose first $k$ statistical moments are (approximately) equal to those of the Haar ensemble $\ensH$ is called to form (approximate) quantum state $k$-designs. These moments of the Haar ensemble $\rho^{(k)}_{\rm H} = \int_{\kpsi \sim \ensH} \dif\mu(\kpsi) \left( \ketbra{\psi}{\psi} \right)^{\otimes k}$ have well-established closed form expressions~\cite{harrow2013churchsymmetricsubspace}, which enable direct numerical comparisons between an ensemble and $\ensH$.

To examine whether ${\cal E'}$ forms quantum state $k$-designs, we thus compare its $k$th moments %
\begin{align}
    \rho^{(k)}_{\cal E'} = \sum_{s} p(s | \rho) \left( \ketbra{\psi_s}{\psi_s}\right)^{\otimes k} 
\end{align}with those of the Haar ensemble $\rho^{(k)}_{\rm H}$, where $\rho = \ketbra{\overline{0}}{\overline{0}}$. Specifically, we shall compute their trace distance
\begin{align}
    \Delta^{(k)} \coloneqq \left\Vert \rho^{(k)}_{\cal E'} - \rho^{(k)}_{\rm H}\right\Vert_1,
\end{align}where $\left\Vert X \right\Vert_1 = \Tr [\sqrt{X^\dagger X}]$ is the trace norm. Such distances $\Delta^{(k)}$ not only indicate how (dis)similar two ensembles are, but they also imply, for example, that any strategy to distinguish these ensembles using $k$ copies of the state has success probability $p_{\rm succ} \leq \frac{1}{2} + \Delta^{(k)}$~\cite{Helstrom1969}, which approaches that of a random guess if $\Delta^{(k)} \to 0$. 

Numerically, we reconstruct the state $\kpsis$ using Eqs.~(\ref{eq:pz0}--\ref{eq:py}), and thus estimate the moments $\rho^{(k)}_{\cal E'}$ by averaging over $R$ sampled syndromes. Using the analytically known $\rho^{(k)}_{\rm H}$ moments~\cite{harrow2013churchsymmetricsubspace}, we then estimate $\Delta^{(k)}$. The $k=1$ moments trivially agree within statistical errors due to the randomly applied $\XL$ in the construction of~$\cal E'$. For $k>1$, our data suggests that 
$\Delta^{(k)}$ decreases with $\phi$, and above the critical angle $\phi_{\rm th}$, approaches a value set by the number $R$ of syndromes averaged over [Fig.~\ref{fig:BornDesigns}]. This value represents the trace distance between the $k$th moments of the true and an empirical Haar ensemble, the latter containing $R$ states. Moreover, in the insets of Fig.~\ref{fig:BornDesigns}, we focus on the data points that allow for a diffusive metal [$L> \ell(\phi)$]; these suggest that $\Delta^{(k)}$ systematically increases (decreases) above (below) threshold. Hence, our data is consistent with $\cal E'$ approaching the Haar ensemble in the diffusive regime. 

\section{Discussion and Outlook}
\label{sec:discOut}

We have studied the measurement-induced randomness of the PLE in surface codes subject to uniform deterministic single-qubit Pauli $X$ rotations, syndrome measurements, and maximum-likelihood Pauli corrections. To shed light on this randomness, using that the states in the PLE are encoded in partially contracted 2D Gaussian fermionic tensor networks, we demonstrated that these are also encoded in the final states of (1+1)D Gaussian quantum circuits. By recasting these circuits as Majorana scattering networks, we have established a one-to-one correspondence between the final states and scattering matrices of a disordered quantum dot. Via this mapping, we have shown that the emergent universal RMT description of this synthetic quantum dot translates into an emergent universality of the PLE.

The PLE that we describe may prove useful for designing schemes to inject non-stabilizer states into the code space using transversal operations. (See also Ref.~\cite{Huang2025,Nobu_transversal_2025} for a related construction, focusing on $\phi < \phi_{\rm th}$.)
The case $\phi=\pi/8$, which lies above the honeycomb SC threshold $\phi_{\rm th}\approx 0.09\pi$, is particularly interesting because the transversal physical rotations are $HTH$, with single-qubit Hadamard $H$ and $T \propto e^{i\frac{\pi}{8} Z}$ gates.  %
These can be implemented using standard fault-tolerant Clifford+$T$ gadgets~\cite{Gottesman_1999,Bravyi_Kitaev_PhysRevA.71.022316,nielsen2010quantum}.
The PLE is a heralded ensemble of states that are generically non-stabilizer, i.e., magic states.
Furthermore, the logical-level operation $D_s$ is known, and the syndrome-resolved angles $(\varphi_s,\theta_s)$ form a highly random heralded ensemble. 
Analogously to the setting discussed in Ref.~\cite{eckstein2025learningtransitionstopologicalsurface}, this may enable shadow-style logical-state tomography using Pauli measurements.
Whether the PLE can be converted into useful magic-state inputs, or otherwise exploited computationally or tomographically, remains an important direction for future work and will require extending the analysis beyond perfect syndrome measurements and ultimately to a circuit-level framework that includes stochastic noise.

We mostly focused on the diffusive regime (which we could access for $\phi \gtrsim \phi_{\rm th}$), where the onset of the universal PLE is expected to be controlled by the finite-size scaling of the bulk conductivity $g(L)$. This is an intuitive expectation that is mathematically further motivated by supersymmetric NL$\sigma$M results that capture the universal (i.e., RMT) regime; however, these results describe systems corresponding to the $N\to 0$ replica limit~\cite{altshuler1986repulsion,efetov1983supersymmetry,Fyodorov_95_PhysRevB.51.13403,efetov1999supersymmetry,mirlin2000statistics,bocquet2000disordered}.  
It is an interesting question whether the emergence of, e.g., conductance moments or distribution consistent with the universal PLE can be established analytically for our $N\to 1$ setting. This may motivate the development of a supersymmetric Born-rule NL$\sigma$M, given the known difficulties in accessing the nonperturbative RMT regime using fermionic replica techniques~\cite{Yurkevich_PhysRevB.60.3955,Kamenev_PhysRevB.60.3944,kamenev1999wigner,verbaarschot1985critique,zirnbauer1999another}.

In the honeycomb SC with larger $\phi$, we see evidence of a crossover to a ballistic regime, akin to when the mean free path exceeds the system size, which seems distinct from the $\phi = \pi/4$ ballistic metal, and which exhibits a closer approach
of the PLE to the Haar ensemble (albeit in a manner less systematic with $L$).
This raises a natural open question of whether there is an appropriate analytical model that can describe this mesoscopic regime, which may be particularly relevant to experiments with moderate-size codes.

It would also be interesting to extend our approach to surface codes with multiple logical qubits, and to settings with more general unitaries, beyond Pauli-$X$ rotations. Since the surface code states are isometric tensor network states, applying local gates followed by measurements will always map to a (1+1)D nonunitary circuit~\cite{mie7}, but generically this will not be Gaussian~\cite{DarmawanTensor,Huang2025,generic1qCohErr,behrends2025surfacecodepaulichannels}. While the description of the PLE in terms of the scattering matrix relies on Gaussian operations through its underlying scattering network, our embedding of the SC logical space in quantum circuit space, and its relation to tensor network descriptions have natural generalizations. In non-Gaussian cases, the entanglement phases of the long-time state of the circuit~\cite{Jian:2022jg,jbfvbb_prr,generic1qCohErr,behrends2025surfacecodepaulichannels} subsume transport properties (e.g., diffusive conduction), and we expect that logarithmic- and volume-law~\cite{generic1qCohErr,bao2024phasesdecodabilitysurfacecode} entanglement phases may also imply convergence to RMT behavior~\cite{wang2026entanglementtransitionstranslationinvarianttensor} and a corresponding universal PLE~\cite{cheng2025}.
Establishing this, and potential links to mesoscopics would be very interesting.

In a similar vein, it would be interesting to generalize beyond the $\mathbb{Z}_2$ surface codes, for instance, to string-net models, which also admit an isometric tensor network representation \cite{soejima2020isometric}, and hence can be mapped to (1+1)D nonunitary circuits. Viewing the transversal unitary evolution of the code as a particular case of wavefunction deformation furthermore suggests studying the PLE for more general states and  deformations~\cite{deformed1,deformed2,deformed3,deformed4,deformed5,deformed6, deformed7, deformed8}.
How do the algebraic properties of these states imprint themselves on the associated circuit, and is there a similar interpretation of the PLE in terms of mesoscopic physics?

The PLE, furthermore, may be viewed as arising from a mixed-state $\rho'=\sum_s p(s|\rho)\ket{\psi_s}\bra{\psi_s}$ obtained from $U\ket{\psi}$ (with $\rho=\ket{\psi}\bra{\psi}$) upon non-selective syndrome measurement. This links the PLE to the classification of mixed-state quantum phases and their relation to QEC~\cite{coser2019classification,bao2023mixedstatetopologicalordererrorfield,Fan_PRXQuantum.5.020343,Rakovszky_PhysRevX.14.041031,Chen_PhysRevLett.132.170602,Chen_PhysRevB.110.125152,Li_PhysRevX.15.011068,Sang_PhysRevLett.134.070403,Sang_PhysRevX.14.031044,Lessa_PRXQuantum.6.010344,Sala_PhysRevB.110.155150,Sa_1lzb-vp1w,negari2024spacetimemarkovlengthdiagnostic,negari2025symmetryenforcesentanglementhigh,negari2026criticalnonequilibriumphasesnoisy,hauser2026strongtoweaksymmetrybreakingopen}. From this perspective, a natural question is how the statistical properties of the PLE, along with the many-body properties of $\ket{\psi_s}$, may help characterise the corresponding mixed-state phases, and what the implications of the approach to the Haar ensemble are~\cite{negari2026quantum_ensemble}.

Finally, the new insights provided by our results could be used to shed light on other measurement-induced phenomena. While most studies of projected ensembles do not explicitly involve QEC codes, it is possible that code-related concepts could offer a complementary perspective on more generic kinds of measurement-induced dynamics. Indeed, the measurement-induced phase transition that arises in monitored quantum circuits can be interpreted as a coding transition \cite{Choi_PhysRevLett.125.030505,gullans2020purification}, suggesting that these ideas could also be explored in that setting.

\textit{Note added.} Upon completion of this manuscript, we became aware of two independent works that discuss how different lattices of topological codes affect the associated symmetry class~\cite{yan2026nonlinearsigmamodelsurface,Yang:2026ktu}. We discussed the relation between these and our work in Sec.~\ref{sec:relation}.

\section*{Acknowledgments} We thank Ehud Altman, Yimu Bao, Michael Gullans, Timothy Hsieh, Chao-Ming Jian, Andreas W.~W.~Ludwig, Amir-Reza Negari, Subhayan Sahu, Simon Trebst, Sagar Vijay, Stephen Yan, Zhou Yang, and Guo-Yi Zhu for helpful discussions and collaboration on related topics. We also thank Chris Self and David Amaro for their feedback on the manuscript. This work was supported by EPSRC PhD studentship 2606484, CPS studentship RS2025/005, UKRI FLF grant MR/Z000297/1 (MB); a Leverhulme Early Career Fellowship and the Newton Trust of the University of Cambridge (JB); Trinity College, Cambridge (MM); EPSRC grant EP/V062654/1 (JB, BB); and in part by the U.S. DOE's Office of Science, Office of Nuclear Physics, InQubator for Quantum Simulation under Award No. DE-SC0020970, and grant NSF PHY-2309135 to the Kavli Institute for Theoretical Physics~(BB). 

Our simulations used resources at the Cambridge Service for Data Driven Discovery operated by the University of Cambridge Research Computing Service (\href{https://www.csd3.cam.ac.uk}{www.csd3.cam.ac.uk}), provided by Dell EMC and Intel using EPSRC Tier-2 funding via grant EP/T022159/1, and STFC DiRAC funding (\href{https://www.dirac.ac.uk}{www.dirac.ac.uk}).

\appendix

\section{Stabilizer graph constraints on the logical ensemble} \label{app:constraints}

Here we review how the parity of the weights of $S^X_v$ stabilizers can constrain $\kpsis$~\cite{Bravyi:2018ea,Venn:2020ge}, as mentioned in Sec.~\ref{sec:LE}. We expand $U $
\begin{align}
    U = \prod_{j=1}^N \exp(i\phi X_j) = 
    \sum_{x \in \{0,1\}^N} c_x i^{| x |} X(x), \label{eq:Ucx}
\end{align}where $|x|$ is the Hamming weight of bitstring $x$, and with Pauli string $X(x) = \prod_j X^{x_j}_j$, and real coefficients $c_x \in \mathbb{R}$. Note that $C_s = X(u_s)$ for a bitstring $u_s$. Consider next
\begin{align}
    D_s = \sum_{x \in \{0,1\}^N} c_x i^{|x|}\Pi_0 X(x \oplus u_s) \Pi_0,
\end{align}where $\oplus$ denotes bitwise addition. Using Eq.~\eqref{eq:Ds}, we find
\begin{align}
    {\cal Z}_{0,s} = \sum_{x \in {\cal X} \oplus u_s} c_x i^{|x|}, \ \ \  {\cal Z}_{1,s} = \sum_{x \in {\cal X} \oplus u_s \oplus \ell} c_x i^{|x|}, \label{eq:ZsBitsX}
\end{align}where $\cal X$ collects all bitstrings compatible $S^X_v$ stabilizers ${\cal X} \coloneqq \{ x \in \{0,1\}^n \ | \ X(x) \in \langle \{ S^X_v \} \rangle \}$, the bitstring $\ell$ gives $X(\ell) = \XL$, and the set ${\cal X} \oplus y = \{ x \oplus y \ | \ x\in {\cal X} \}$. Hence, we recast Eq.~\eqref{eq:ZsBitsX} as
\begin{align}
    {\cal Z}_{0,s} &= \sum_{x \in {\cal X}} c_{x\oplus u_s} \ i^{|x\oplus u_s|}, \label{eq:Z0bits} \\ {\cal Z}_{1,s} &= \sum_{x \in {\cal X}} c_{x \oplus u_s \oplus \ell} \ i^{|x \oplus u_s \oplus \ell|}. \label{eq:Z1bits}
\end{align}

The parity dependence is finally apparent from Eqs.~\eqref{eq:Z0bits} and~\eqref{eq:Z1bits}. If some $S^X_v$ stabilizers have odd weight (i.e., $|x|$ is odd for some $x \in {\cal X}$), the relative phase between ${\cal Z}_{0,s}$ and ${\cal Z}_{1,s}$ can take an arbitrary value. Conversely, if all $S^X_v$ stabilizers have even weight ($|x|$ even $\forall x \in {\cal X}$), we have $\arg(\mathcal{Z}_{0,s}/\mathcal{Z}_{1,s}) = \varphi_s \in  \{\varphi, \varphi+ \pi\} \ \forall s$ with $\varphi = 0$ for $|\ell|$ even, and $\varphi = \pi/2$ for $|\ell|$ odd. To see this, one can use $c_x \in \mathbb{R}$ and the relation $|x \oplus y|  \textrm{ mod }2 = (|x| + |y|) \textrm{ mod }2$.

\section{Quantifying the randomness of the logical ensemble via the Wasserstein-1 distance}
\label{app:LErandomness}

In this Appendix, we discuss how to compare the logical ensemble $\ens$ [Sec.~\ref{sec:LE}] with a constrained Haar ensemble $\ensH^*$ using the Wasserstein-1 distance.

Statistical distances between $\ens$ and $\ensH^*$, viewed as probability distributions on the upper Bloch hemisphere, are a natural way to compare $\ens$ with $\ensH^*$~\cite{Villani2009}. In choosing such a distance, one crucially needs to accommodate for $\ens$ and $\ensH^*$ being discrete and continuous, respectively. In particular, although the total variation distance and the Kullback-Leibler divergence are appropriate choices of statistical distances in many probability-theoretic settings, these are not amenable here because they attain maximal values when comparing a discrete with a continuous distribution. Instead, the Wasserstein-1 distance has been recently proposed as a suitable metric to compare discrete and continuous ensembles of quantum states~\cite{Bejan:2025ch}.

The Wasserstein-1 distance is formally defined, for two probability measures $\mu$ and $\nu$ on a space $\mathcal{M}$ endowed with a metric $d$, as~\cite{Villani2009}
\begin{align}
    W_1(\mu, \nu) = \inf_{\gamma \in \Pi(\mu, \nu)} \left(\mathbbm{E}_{(x,y) \sim \gamma}  d(x,y) \right)
\end{align}
where the set $\Pi(\mu, \nu)$ collects all couplings of $\mu$ with $\nu$, i.e., probability measures $\gamma$ over $\mathcal{M}\times \mathcal{M} \ni (x,y)$ such that the marginals on each copy of $\mathcal{M}$ are $\mu$ and $\nu$, respectively. This highly general definition is related to the mathematical problem of optimal transport, in which the optimal distribution $\gamma$ represents the most efficient way to transport probability mass between different regions of $\mathcal{M}$, with the distance that each mass element is moved quantifying the efficiency~\cite{Villani2009}. For our purpose, it will be more convenient to use the Kantorovich-Rubinstein duality to represent the Wasserstein metric as~\cite{Kantorovich1958}
\begin{align}
    W_1(\mu, \nu) = \sup_{\substack{f : \mathcal{M} \rightarrow \mathbbm{R}\\ \textrm{Lip}(f) \leq 1}} \mathbbm{E}_{x \sim \mu}[f(x)] - \mathbbm{E}_{x \sim \nu}[f(x)], \label{eq:WassersteinDefMain}
\end{align}
with the expectation of $f$ over the distribution $\mu$ denoted by $\mathbbm{E}_{x \sim \mu} [f(x)]$, and analogously for $\nu$, and Lipschitz constant ${\rm Lip} (f) = \sup_x \Vert \nabla f (x) \Vert_2$, with Frobenius norm $\Vert X \Vert_2 = \sqrt{\Tr[X^\dagger X]}$. To probe the discrepancies between the distributions $\mu$ and $\nu$, one uses the optimal test function $f(x)$, which is chosen over all sufficiently smooth functions to maximize the discrepancy. Here, these smooth functions are $c$-Lipschitz with $c \leq 1$
\begin{align}
    |f(x_1) - f(x_2)| &\leq c\, d(x_1, x_2) & \forall x_1, x_2 \in \mathcal{M},
\end{align}
The specific distributions that we compare here are $\mu = \ens$ and $\nu = \ensH^{(*)}$ [Sec.~\ref{sec:LE}], and the space ${\cal M}$ is (a subspace of) the upper hemisphere of the logical Bloch sphere.
 
Computing the Wasserstein-1 distance between $\ens$  and $\ensH^*$ is a tremendous task due to the need to optimize over functions $f$ in Eq.~\eqref{eq:WassersteinDefMain}.
While this $W_1(\ens, \ensH^*)$ may be numerically approximated using linear programming~\cite{cOptTrans}, here we instead consider the distribution of a specific scalar quantity $\vartheta \colon {\cal M} \to X \subseteq \mathbbm{R}$. 
This enables us to exactly compute the Wasserstein-1 distance of the two univariate distributions induced by $\ens$ and $\ensH^*$, and which can be represented by their density functions $q(\vartheta)$ and  $q'(\vartheta)$ over the domain $X$, respectively. Eq.~\eqref{eq:WassersteinDefMain} then simplifies significantly to~\cite{w1CDFs} 
\begin{align}
    W^{(\vartheta)}_1 \equiv W_1\big(q(\vartheta),q'(\vartheta)\big) = \int_{\vartheta \in X} \dif \vartheta |Q(\vartheta) - Q'(\vartheta)|,
    \label{eq:W11D}
\end{align}
with $Q(\vartheta)$ and $Q'(\vartheta)$ being the cumulative distribution functions (CDFs) of $q(\vartheta)$ and $q'(\vartheta)$, respectively, i.e., one takes the unsigned area between the CDFs. This will allow us to numerically compute $W^{(\vartheta)}_1$ in certain instances.

These scalar functions $\vartheta$ encompass various physically-relevant quantities, such as the subsystem entanglement entropies, purity, $n$-point correlation functions (see, for example, Ref.~\cite{Bejan:2025ch}), as well as QEC-relevant quantities, including the ratio $\min_q [p_{q,s}/p(s)]$ and the conductance $G$ of a quantum dot associated with the code of interest [Sec.~\ref{sec:quantum_dot}]. In Sec.~\ref{sec:qdotNum}, we focus on the distribution of the conductance $\vartheta = G$. Thus, by considering $W^{(G)}_1$, we shall compare the empirical distribution built from $G_s \equiv G(\kpsis)$ with that of $G(\ket{\phi})$, where $\kpsis$ and $\ket{\phi}$ are distributed according to $\ens $ and $\ensH^*$, respectively. Using estimates of such Wasserstein-1 distances, we assess the randomness of the PLE in Sec.~\ref{sec:W1Born}.

\section{Sampling algorithm}
\label{app:sampling}

For coherent errors, sampling of syndromes is nontrivial: error strings $C_s$ that define $\{ \eta_j\}$ cannot be sampled independently on each site, as it is the case for Pauli noise, but need to be drawn from a correlated probability distribution.
The algorithm described in Ref.~\onlinecite{Bravyi:2018ea} solves this problem via a mapping of qubits to fermions~\cite{Kitaev_2006} and thereby introduces a Gaussian tensor network, which can be generalized to other planar graphs~\cite{Venn:2020ge}.
Here we describe a variation of the algorithm that ties in more closely with the conventions used in this paper, and can be generalized to more generic error channels~\cite{generic1qCohErr}.

Starting point is Eq.~\eqref{eq:absolute_value}, which gives the probability of an error string $C_s \XL^q$.
The quantum circuit $\calM$ consists of $N$ tensors $\hat{T}_\eta$, where $N$ is the number of qubits.
As described in the main text, the $\hat{T}_\eta$ can be reshaped as two-qubit gates $\hat{V}^{(l,m)} \to \hat{V}_j^{(\eta_j)}$ alternating with layers $\hat{H}$ and $\hat{H}'$, where we introduced a label $j=1,\dots ,n$ for convenience, assuming gate ordering $\calM = V_N^{(\eta_n)} \dots V_2^{(\eta_2)} V_1^{(\eta_1)}$, and explicitly retained the labels $\eta_j$ that define the error string $C_s = \prod_j X_j^{(1-\eta_j)/2}$.
Using this notation, the probability of $C_s$ is
\begin{equation}
 p_\eta = \braket{ \phi^\text{(i)} | [\hat{V}_1^{(\eta_1)}]^\dagger \cdots [\hat{V}_n^{(\eta_n)}]^\dagger \omega_{\phi}^{(n)} \hat{V}_n^{(\eta_n)} \cdots \hat{V}_1^{(\eta_1)} | \phi^\text{(i)} },
\end{equation}
where $\omega_{\phi}^{(n)} \equiv \rho_0^{(f)}$ is the density matrix of the final state.

Now suppose the signs $\eta_1, \dots, \eta_{n-1}$ are already sampled, and that we want to sample $\eta_n$. Its conditional probability $p(\eta_n|\eta_{n-1}, \dots, \eta_{1} )$ equals $p_\eta$ divided by the marginal probability of the first $n-1$ qubits
\begin{equation}
 p_{n-1} (\eta_{n-1},\dots ,\eta_1) = \braket{ \phi_{n-1}^{\{\eta\}} | \omega_\phi^{(n-1)} | \phi_{n-1}^{\{\eta\}} },
\end{equation}
where $\ket{\phi_{n-1}^{\{\eta\}}} = \hat{V}_{n-1}^{(\eta_{n-1})} \dots \hat{V}_1^{(\eta_1)} \ket{ \phi^\text{(i)} }$ is the initial state $\ket{ \phi^\text{(i)} }$ evolved by the gate sequence up to the penultimate gate and $\omega_\phi^{(n-1)} = \sum_{\eta_n} [\hat{V}_n^{(\eta_n)}]^\dagger \omega_\phi^{(n)} \hat{V}_n^{(\eta_n)}$ is the sum over the two choices for $\eta_n$.
Generally, the conditional probability of the $j^\text{th}$ qubit is the ratio of marginal probabilities $p(\eta_j|\eta_{j-1},\dots, \eta_1) = p_j(\eta_{j},\dots, \eta_1)/p_{j-1} (\eta_{j-1},\dots, \eta_1)$ for all $j=1,\dots, n$, with~\cite{generic1qCohErr}
\begin{equation}
 p_{j} (\eta_{j},\dots, \eta_1) = \braket{ \phi_{j}^{\{\eta\}} | \omega_\phi^{(j)} | \phi_{j}^{\{\eta\}} } ,
 \label{eq:pj_marginals}
\end{equation}
where $\omega_\phi^{(j)}$ can be iteratively obtained from $\omega_{\phi}^{(n)}$.

We have used in the main text that $\ketbra{\phi_{j}^{\{\eta\}} }{\phi_{j}^{\{\eta\}} }$, when projected onto a fixed-parity sector, is Gaussian.
Here we show that $\omega_\phi^{(j)}$ is also Gaussian, which means we can efficiently sample error strings using this approach.
To this end, consider the gate $\hat{V}_{j}^{(\eta_{j})} = e^{K + i J \gamma_{m_j} \gamma_{m_j+1}}$, where $(m_j,m_j+1)$ denotes the pair of Majorana indices that $\hat{V}_{j}^{(\eta_{j})}$ acts on.
The transformation $\omega_\phi^{(j)} \mapsto \omega_\phi^{(j-1)}$ simplifies to
\begin{align}
 \omega_\phi^{(j-1)}
 &= \Phi_j [\omega_\phi^{(j)}] =  \sum_{\eta_{j}} [\hat{V}_{j}^{(\eta_{j})}]^\dagger \omega_\phi^{(j)} \hat{V}_{j}^{(\eta_{j})} \\
 &= \frac{1}{2} \left( \omega_\phi^{(j)} + \gamma_{m_j} \gamma_{m_j+1} \omega_\phi^{(j)} \gamma_{m_j+1} \gamma_{m_j} \right) , \nonumber
\end{align}
where we have defined the channel $\Phi_j [\dots]$ and used that $e^{2\Re K} \cosh J \cosh J^* = e^{2\Re K} \sinh J \sinh J^* = 1/4$, where $K,J$ are defined in Eq.~\eqref{eq:rbim_coefficients}.
The channel $\Phi_j$ is a Gaussian transformation: following Ref.~\onlinecite{Bravyi:2005jh}, we consider the operator $\rho_{\Phi_j} = (\Phi_j \otimes_f \id) [\rho_I]$ dual to $\Phi_j$, where $\otimes_f$ is the fermionic tensor product~\cite{Bravyi:2005jh}, and where the density matrix $\rho_I = \frac{1}{4^{M}} \prod_{j=1}^{2M} \left( 1 + i \gamma_j \gamma_{j+2M} \right)$ is defined in a doubled Hilbert space. Since
\begin{equation}
\rho_{\Phi_j} =  \Phi_j [\rho_I] = \frac{1}{4^{M}} \prod_{j \neq \{m_j,m_j+1\}} \left( 1 + i \gamma_j \gamma_{j+2M} \right)
\end{equation}
remains Gaussian (albeit not pure), $\Phi_j$ must be a Gaussian map.
The effect on the previous state $\omega_\phi^{(j)}$ can be understood as effectively removing the pair $(m_j,m_j+1)$ from $\omega_\phi^{(j)}$.

After applying all gates $\hat{V}_{j}^{(\eta_{j})}$ within one layer of the quantum circuit, we need to apply either $\hat{H}$ or $\hat{H}'$. Since these consists only of projectors $\frac{1}{2} (1+ i \gamma_{2m}\gamma_{2m+1})$, they are also Gaussian.
Hence, all $p_{j} (\eta_{j}\dots \eta_1)$ can be efficiently computed in polynomial time.

Since each gate $\Phi_j$ effectively removes a pair of Majoranas, the initial $\omega_\phi^{(N)}$ will have Majoranas gradually removed until only a term $\propto (1+i \gamma_1 \gamma_{2(M+1)})$ remains; cf.\ Eq.~\eqref{eq:final_states}.
Th
This implies that the marginal probabilities [Eq.~\eqref{eq:pj_marginals}] will gradually involve fewer Majoranas in $\omega_\phi^{(j)}$ and thereby slowly evolve (but never reach) towards the identity operator. (Each $\hat{H}$ and $\hat{H'}$ layer again changes this to a projector with elements $\frac{1}{2} (1+ i \gamma_{2m}\gamma_{2m+1})$ and hence resets this tendency towards the identity $\id$.)
This property is basis for the ``Born-rule'' argument in the main text: it is largely the norm of the state $\ket{\phi_{j}^{\{\eta\}}}$ for both values of $\eta_j$ that determines the probability for each $\eta_j =\pm 1$~\cite{Yang:2026ktu}.

\section{Fermionic linear optics representation}
\label{app:flo}

Here we provide more details on the Gaussian tensor network used in the main text and show how to contract it in polynomial time~\cite{Bravyi:2005jh}.
Key for the polynomial-time contraction is that all information about intermediate contraction steps, described by the state $\ket{\phi_{j}^{\{\eta\}}} = \hat{V}_{j}^{(\eta_{j})} \cdots \hat{V}_1^{(\eta_1)} \ket{ \phi^\text{(i)} }$ (using the notation from App.~\ref{app:sampling}), is fully contained in its norm and the correlation matrix $\Gamma^{(j)}$ with elements
\begin{equation}
 \Gamma_{kl}^{(j)} = \frac{i}{2} \frac{\braket{\phi_j^{\{\eta\}} | [\gamma_k,\gamma_l] |\phi_j^{\{\eta\}}}}{\braket{\phi_j^{\{\eta\}} |\phi_j^{\{\eta\}}}}.
\end{equation}

We now compute the fermionic linear optics representation of the transformation via $\hat{V}_j^{(\eta_j)}$~\cite{Bravyi:2005jh,jbfvbb_prr} that is necessary to describe the evolution of $\Gamma^{(j-1)} \mapsto \Gamma^{(j)}$ during the contraction process.
The operator $\hat{V}_j^{(\eta_j)} =e^{K + i\eta_j J \gamma_{m_j}\gamma_{m_j}}$ transforms the correlation matrix $\Gamma^{(j-1)}$ of a Gaussian state as~\cite{Bravyi:2005jh}
\begin{align}
 \Gamma^{(j-1)} \mapsto  \Gamma^{(j)} = B (1 - \Gamma^{(j-1)} A)^{-1} \Gamma^{(j-1)} B^T + A
\end{align}
where $B$ equals the identity matrix apart from the $2 \times 2$ sector spanned by the indices $(m_j,m_j+1)$, where it equals $B_{m_j,m_j} = B_{m_j+1,m_j+1} = 0$ and $B_{m_j,m_j+1} = - B_{m_j+1,m_j}  = \sin (2\phi)$.
The elements of $A$ equal zero apart from the $2 \times 2$ sector spanned by the indices $(m_j,m_j+1)$ with $A_{m_j,m_j} = A_{m_j+1,m_j+1} = 0$ and $A_{m_j,m_j+1} = - A_{m_j+1,m+1} = \cos (2\phi)$.
The norm of the new density matrix is~\cite{Bravyi:2005jh}
\begin{equation}
 \braket{\phi_{j-1}^{\{\eta\}} | [\hat{V}_j^{(\eta_j)}]^\dagger \hat{V}_j^{(\eta_j)} |\phi_{j-1}^{\{\eta\}}}
 = \frac{1 + \eta_j \cos \phi \Gamma^{(j-1)}_{m_j,m_j+1}}{2}%
\end{equation}

\section{Details on the scattering matrix of the quantum dot} \label{app:dot_scatt_mat}

In this Appendix, we explicitly construct the scattering matrix of the synthetic quantum dot $S[\tilM]$ and show that this is isomorphic to the correlation matrix of the final 2-qubit state $\ket{\phi_s}$.

\subsection{Construction from the effective Gaussian tensor} \label{app:SfromM}

Here, we derive in more detail the scattering matrix of the quantum dot discussed in Sec.~\ref{sec:qdot_scatt} [see Eq.~\eqref{eq:sdot}]. By considering the single-particle action of $\tilde{\cal M}_s$ [Eq.~\eqref{eq:tildeM}], which we reproduce here for convenience 
\begin{align}
    \tilde{{\cal M}}_s = \frac{ \cos\frac{\theta_s}{2} (1+i\gamma_1 \gamma_2) + \sin\frac{\theta_s}{2} e^{i\varphi_s}  (1-i\gamma_1 \gamma_2) }{\sqrt{2}},
\end{align}we first construct the transfer matrix $t_p[\tilde{\cal M}_s]$ in the particle sector via~\cite{Jian:2022jg} (see also Sec.~\ref{sec:scatt_net})
\begin{align}
    \tilde{\cal M}_s \boldsymbol{\gamma} \tilde{\cal M}_s^{-1} = \boldsymbol{\gamma}^T t_p[\tilde{\cal M}_s],
\end{align}with $\boldsymbol{\gamma}= (\gamma_1, \gamma_2)^T$ and $t_p[\tilde{\cal M}_s] = \begin{psmallmatrix}
    A_s & B_s \\ -B_s & A_s
\end{psmallmatrix}$ where
\begin{align}
    A_s &=  \frac{1}{\sin \theta_s} \left[\cos^2\left( \frac{\theta_s}{2}\right) e^{-i\varphi_s} + \sin^2\left( \frac{\theta_s}{2}\right) e^{i\varphi_s}\right], \\
    B_s &= \frac{i}{\sin \theta_s}\left[\cos^2\left( \frac{\theta_s}{2}\right) e^{-i\varphi_s} - \sin^2\left( \frac{\theta_s}{2}\right) e^{i\varphi_s}\right].
\end{align}Note that since $\tilde{\cal M}_s = a\id + b X$ for Pauli $X$ and $a,b \in \mathds{C}$, its inverse is $\tilde{\cal M}_s^{-1} = \frac{a\id - b X}{a^2-b^2}$, which exists for $a^2 \neq b^2$. Using the time-reversed $t_h[\tilde{\cal M}_s] = t^*_p[\tilde{\cal M}_s]$~\cite{Jian:2022jg}, we form the doubled transfer matrix $\mathsf{t}'[\tilde{\cal M}_s] = 
    t_p[\tilde{\cal M}_s] \oplus t^*_p[\tilde{\cal M}_s]$. Then, working in the rotated basis 
    $\mathsf{t}[\tilde{\cal M}_s] = U \mathsf{t}'[\tilde{\cal M}_s] U^\dagger,$
with $U = e^{i \frac{\pi}{4}\Sigma_x}$ yields $\mathsf{t}$ satisfying $\mathsf{t}^{-1} = \Sigma_z \mathsf{t}^\dagger \Sigma_z$. Finally, we can relate the blocks of $\mathsf{t}[\tilde{\cal M}_s]$ to transmission and reflection matrices of a scattering matrix using
\begin{align}
    \hspace{-1em} S[\tilde{\cal M}_s] \coloneqq \begin{pmatrix}
        r_s & t'_s \\ t_s & r'_s
    \end{pmatrix}, \  \mathsf{t} [\tilde{\cal M}_s] \equiv
\begin{pmatrix}
 {t_s^\dagger }^{-1} & r'_s {t'_s}^{-1} \\
 -{t'_s}^{-1} r_s & {t'_s}^{-1}
\end{pmatrix}.
\end{align}%
Hence, we find the transmission matrices
\begin{align}
    t_s = t'^T_s = \sin \theta_s \begin{psmallmatrix}
        \cos \varphi_s & \sin \varphi_s \\
        -\sin \varphi_s & \cos \varphi_s
    \end{psmallmatrix} \label{eq:ts}
\end{align}and reflection matrices
\begin{align}
    r_s = r'_s = \cos \theta_s \begin{psmallmatrix}
        1 & 0 \\ 0 & 1
    \end{psmallmatrix}.\label{eq:rs}
\end{align}

\subsection{Isomorphism with the final state correlation matrix}
\label{app:isomorph}

Since $\ket{\phi_s}$ [Eq.~\eqref{eq:partial_contr}] is an unnormalized Gaussian state, it is instructive to consider its correlation matrix. Again switching to Majoranas via $\chi_{2j-1} = (\prod_{k<j} \tau_k^x) \tau_j^z$, $\chi_{2j} = (\prod_{k<j} \tau_k^x) \tau^y_j$ for $j=1,\dots,4$, the corresponding correlation matrix $\Gamma_{s,jk}^{(\chi)} = \frac{i}{2} \braket{\phi_s| [\chi_j,\chi_k] | \phi_s }/\braket{\phi_s|\phi_s}$ reads
\begin{align}
    \Gamma^{(\chi)}_s = \begin{psmallmatrix}
 0 & \cos \varphi_s \sin \theta_s &-\sin \varphi_s \sin \theta_s & \cos \theta_s \\
-\cos \varphi_s \sin \theta_s & 0 & \cos \theta_s & \sin \varphi_s \sin \theta_s \\
 \sin \varphi_s \sin \theta_s &-\cos \theta_s & 0 & \cos \varphi_s \sin \theta_s \\
-\cos \theta_s &-\sin \varphi_s \sin \theta_s &-\cos \varphi_s \sin \theta_s & 0 \end{psmallmatrix}.
\end{align}
Two limiting cases are transparent in this formulation: For no errors ($\phi = 0$), we have $\theta_s=0$.\footnote{Assuming maximum likelihood decoding, which for no errors implies $|\mathcal{Z}_{0,s}|^2 = p(s)$ and $|\mathcal{Z}_{1,s}|^2 = p(s)$.} For coherent rotations with $\phi = \pi/4$, we have $\theta_s =\pi/2$ and a fixed phase $\varphi_s \in \{\pi/2,3\pi/2\}$. Furthermore, for the limit to a class D model such that $\varphi_s \in \{ 0, \pi \}$ or $\varphi_s \in \{ \pi/2, 3\pi/2\}$ (see App.~\ref{app:constraints}), one can readily see that $\Gamma^{(\chi)}_s$ decouples into two $2 \times2$ correlation matrices; this showcases that the doubling of fermionic modes is not necessary in class D, as discussed in Sec.~\ref{sec:symm_lattices}.

\begin{figure}[tp]
\centering
\includegraphics[width=\linewidth]{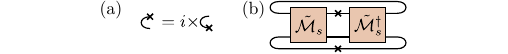}
\caption{(a) Majorana mode (black cross) acting on an empty fermionic state $\ket{0^\text{F}}_m$ that satisfies $\chi_{2m} \ket{0^\text{F}}_m = i\chi_{2m-1} \ket{0^\text{F}}_m$, meaning a factor of $i$ needs to be included when swapping Majorana modes as indicated. (b) The final-state correlation matrix $\Gamma$ of the state $\ket{\phi_s}$ is related to the correlation of the operator $\tilM$ by the process shown in panel (a) and hence to the scattering matrix $S[\tilM]$.}
\label{fig:bending}
\end{figure}

We now interpret the correlation matrix as the correlation matrix of the Gaussian operator $\tilM$ acting on only one fermion (i.e., two Majoranas), cf. also Ref.~\onlinecite{Jian:2022jg}. We visualize this in Fig.~\ref{fig:bending}.
First note that $\chi_1 \ket{0^{\rm F}}_1 = i \chi_2 \ket{0^{\rm F}}_1$ and $\chi_{4} \ket{0^{\rm F}}_{2} = -i \chi_{3} \ket{0^{\rm F}}_{2}$, where $\ket{0^{\rm F}}_m$ denotes an unoccupied fermionic mode on site $m$, equivalent to $\ket{+}_m$ in the $\tau_m^\mu$ transfer matrix space.
Now using the parity-conserving transformation $\chi_1 \to i \tilde{\chi}_1^L$, $\chi_2 \to \tilde{\chi}_1^R$, $\chi_3 \to \tilde{\chi}_2^R$, and $\chi_4 \to -i\tilde{\chi}^L_2$, where the imaginary units arise from the aforementioned back-bending of modes via the unoccupied $\ket{0^{\rm F}}_{m}$ [also cf. Fig.~\ref{fig:bending}(a)], we identify the blocks of the newly defined transformed correlation matrix~\footnote{Note that $\Gamma^{(\tilde{\chi})}$ is complex since not all the operators $\tilde{\chi}_j$ are Hermitian.}
\begin{align}
{\Gamma^{(\tilde{\chi})}_s}
= \begin{pmatrix} \Gamma^{LL}_s & -i\Gamma^{LR}_s \\ i\Gamma^{RL}_s & \Gamma^{RR}_s \end{pmatrix}
\label{eq:correlations_chi_prime}
\end{align}
with the blocks
\begin{align}
\Gamma^{RR}_s &= \Gamma^{LL}_s  = \cos \theta_s e^{i(\pi/2) \tau^y} , \label{eq:Grefl} \\
\Gamma^{RL}_s &= [\Gamma^{LR}_s]^T = \sin \theta_s e^{i \varphi_s \tau^y} \label{eq:Gtrans} .
\end{align}and the Pauli matrix $\tau^y$.
These blocks also correspond to 
\begin{align}
\Gamma_{s,ij}^{LL} &= \frac{i}{2} \Tr \left( [\chi_i, \chi_j] \tilM \tilM^\dagger \right) , \\
\Gamma_{s,ij}^{RR} &= \frac{i}{2} \Tr\left( [\chi_i, \chi_j] \tilM^\dagger \tilM \right), \\
\Gamma_{s,ij}^{RL} &= \Tr\left( \chi_i \tilM \chi_j  \tilM^\dagger \right) , \\
\Gamma_{s,ij}^{LR} &= \Tr\left( \chi_i \tilM^\dagger \chi_j \tilM \right),
\end{align}highlighting that they are another way to represent the operator $\tilM$. This correspondence goes further as we note that $\Gamma^{LL}_s = \Gamma^{RR}_s = r_s = r'_s $ and $t_s = t'^T_s = \Gamma^{RL}_s = [\Gamma^{LR}_s]^T$, cf. Eqs.~\eqref{eq:ts}, \eqref{eq:rs}, \eqref{eq:Grefl}, and \eqref{eq:Gtrans}. Hence, by identifying these blocks, we readily find that the covariance matrix $\Gamma^{(\chi)}_s = \Gamma(\ket{\phi_s})$ is isomorphic to the scattering matrix $S[\tilM]$, as claimed in Eq.~\eqref{eq:SisG}.

\section{Additional class DIII Wasserstein distance numerics} \label{app:addWDIII}
\begin{figure}[tp]
    \centering
    \includegraphics[width=\linewidth]{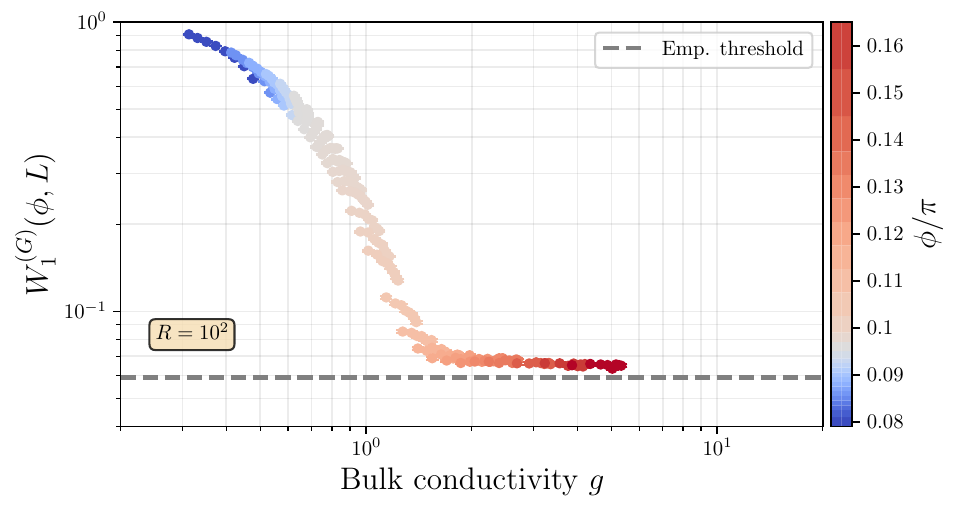}
    \caption{Wasserstein-1 distance $W^{(G)}_1(\phi,L)$ for the quantum dot conductance $G$ between the logical and hemisphere-Haar ensembles vs bulk conductivity $g$  for the honeycomb surface code with rotation angle $\phi$. Data averaged over $10^5$ to $10^6$ syndrome realizations for length and width $21\leq L = M \leq 91$, split into $R=10^2$ syndromes per empirical CDF. Error bars show SE. Above the error threshold $\phi > \phi_{\rm th} \approx 0.089\pi$, $W^{(G)}_1(\phi,L)$ decreases and then approaches an $R$-dependent statistical threshold value (horizontal dashed gray line).}
    \label{fig:BornW1_1e2}
\end{figure}

In this Appendix, we discuss further the dependence of the Wasserstein distance $W_1^{(G)}(\phi,L)$ on the number $R$ of syndromes per empirical CDF in the metallic phase of the honeycomb surface code. As discussed in Sec.~\ref{sec:qdotNum} [see also Fig.~\ref{fig:qdot}(c)], we expect that $W_1^{(G)}(\phi>\phi_{\rm th},L\to \infty)$ approaches an $R$-dependent empirical threshold $\propto 1/\sqrt{R}$. In Sec.~\ref{sec:qdotNum} [Fig.~\ref{fig:qdot}(c)], we considered a large $R=10^4$ for which the saturation to the empirical threshold is beyond the reach of numerically-accessible system sizes $L$. Here, we instead examine such saturation by considering $R=10^2$ syndromes per CDF, thereby yielding a higher empirical threshold.  While full saturation is not reached by the data shown in Fig.~\ref{fig:BornW1_1e2}, $W_1^{(G)}(\phi,L)$ seems to approach the empirical threshold slowly but more closely.%

\section{Partial Pauli twirl class DIII numerics} \label{app:TwirlNumerics}

In this Appendix, we present numerical results for a simplified model of the surface code under coherent rotations: the surface code under ``partial Pauli twirling'', as discussed in Ref.~\onlinecite{jbfvbb_prr}. The key distinction between these two models is how the syndromes, and thus realizations of the corresponding network model, are sampled. Apart from this, they are qualitatively similar: For example, both have a finite error threshold but these differ quantitatively. Focusing on the honeycomb surface code, both of their corresponding scattering networks, and thus NL$\sigma$M, are in symmetry class DIII; however, these correspond to replica limits $N\to 1$ (coherent rotations) and $N\to 0$ (partial Pauli twirl)~\cite{Jian:2022jg, fvjbbb_prl,jbfvbb_prr}; see also Sec.~\ref{sec:implications} and Ref.~\cite{yan2026nonlinearsigmamodelsurface, Yang:2026ktu}. Therefore, their quantum dots are expected to have similar phenomenology since, above the error threshold, their internal fabrics (network models) are metals in class DIII. In the remainder of this Appendix, we spell out the model and then present numerics for the error threshold, insulator-to-metal transition, and emergent RMT behavior of the quantum dot.

\begin{figure}[tp]
    \centering
    \includegraphics[width=\linewidth]{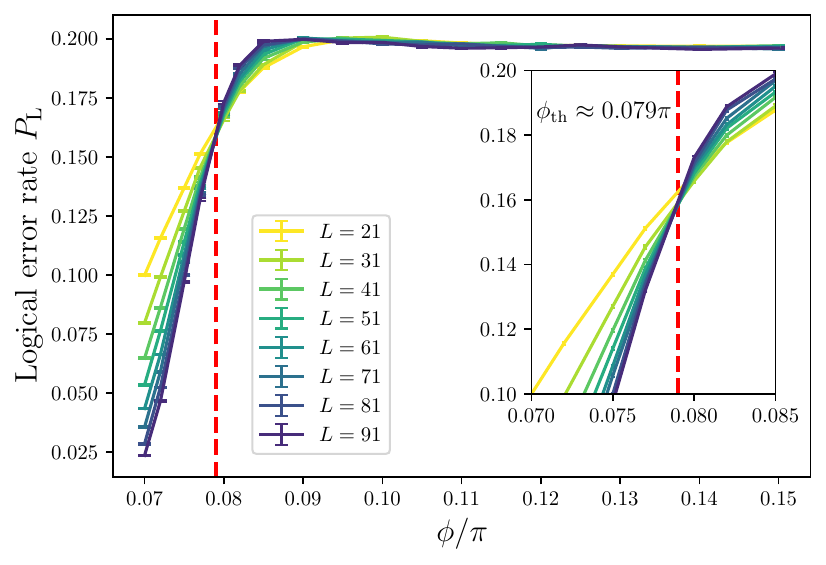}
    \caption{
    Logical error rate $P_{\rm L}$ vs coherent error angle $\phi$ for the honeycomb surface code under partial Pauli twirl with length and width $21\leq L = M \leq 91$. $P_{\rm L}$ decays with $L$ below the error threshold $\phi_{\rm th}$ (vertical dashed red line). The inset shows data closer to the QEC transition. For $\phi > \phi_{\rm th}$, $P_{\rm L}$ initially increases with $L$, after which it becomes weakly dependent on $L$ for the accessible system sizes. Error bars show SE. Data averaged over $10^5$ to $10^6$ syndromes.}
    \label{fig:twirl_err_rate}
\end{figure}

The key difference between partial Pauli twirl and coherent errors is how the couplings $\eta_{vv'}$ in the complex RBIM Hamiltonian are sampled, cf. Eq.~\eqref{eq:partfn}. For coherent errors, $\eta^{(q,s)}_{vv'}$ are sampled according to the Born probability $p(s|\rho)$ %
for syndrome $s$ [see Sec.~\ref{sec:LESC}]. For partial Pauli twirl, $\eta^{(q,s)}_{vv'} \equiv \eta^{(q)}_{vv'}$ are sampled as $-1$ with probability $p=\sin^2 \phi$ and $+1$ with probability $1-p = \cos^2 \phi$, leading to the partition functions and RBIM Hamiltonians
\begin{align}
    \calZ' = \sum_{\{\sigma_v\}} e^{n K + H'_{q,s}} , & & 
 H'_{q,s} = J \sum_{\langle v,v'\rangle} \eta_{vv'}^{(q)} \sigma_v \sigma_{v'},\label{eq:partfnTwirl}
\end{align}with the same complex couplings $K = \frac{1}{2} \ln (\cos \phi \sin \phi)+ i \frac{\pi}{4}$ and $
 J =-\frac{1}{2} \ln (\tan \phi ) - i\frac{\pi}{4}$.
Notably, the partition function $\calZ'$ captures the coherent sums over $X$-strings contributions, which is an essential feature differentiating coherent and incoherent errors.
Before presenting our numerics, we note that how vortex realizations ($\eta_{vv'}=-1$) are sampled does not affect the network model's symmetry class; however, it changes the replica limit of the NL$\sigma$M, and it may change the disorder~averages.

\begin{figure}[t]
    \centering
    \includegraphics[width=\linewidth]{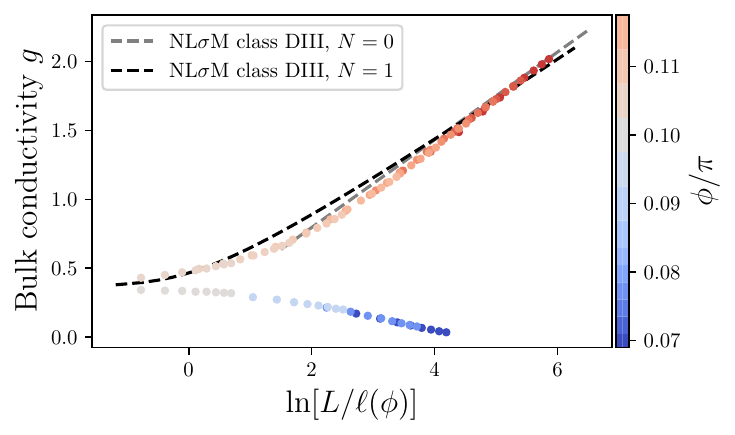}
    \caption{
    Bulk conductivity $g$ for the coherent rotations network model of the honeycomb surface code under partial Pauli twirl with length and width $21\leq L = M \leq 91$. Data averaged over $10^5$ to $10^6$ syndrome realizations. Error bars (SE) are imperceptible. $g$ scales with a $\phi$-dependent length~$\ell(\phi)$. For the insulator $\phi < \phi_{\rm th}$, $g(L/\ell(\phi))$ decreases with $L$, with $\phi_{\rm th} = (0.079 \pm 0.002)\pi$. For the metal $\phi > \phi_{\rm th}$, $g$ increases and, for large $L$, is consistent with the scaling $g \propto \pi^{-1} \ln[L/\ell(\phi)]$ of a nonlinear sigma model in symmetry class DIII with replica limit $N\to 0$  (dashed gray line), see also~\cite{yan2026nonlinearsigmamodelsurface, Yang:2026ktu}. The dashed black line shows the $N\to 1$ limit of the NL$\sigma$M $g \propto (2\pi)^{-1} \ln[L/\ell(\phi)]$.
    }
    \label{fig:twirl_bulk_g}
    \vspace{-\baselineskip}
\end{figure}

\begin{figure*}[htp]
    \centering
    \includegraphics[width=\textwidth]{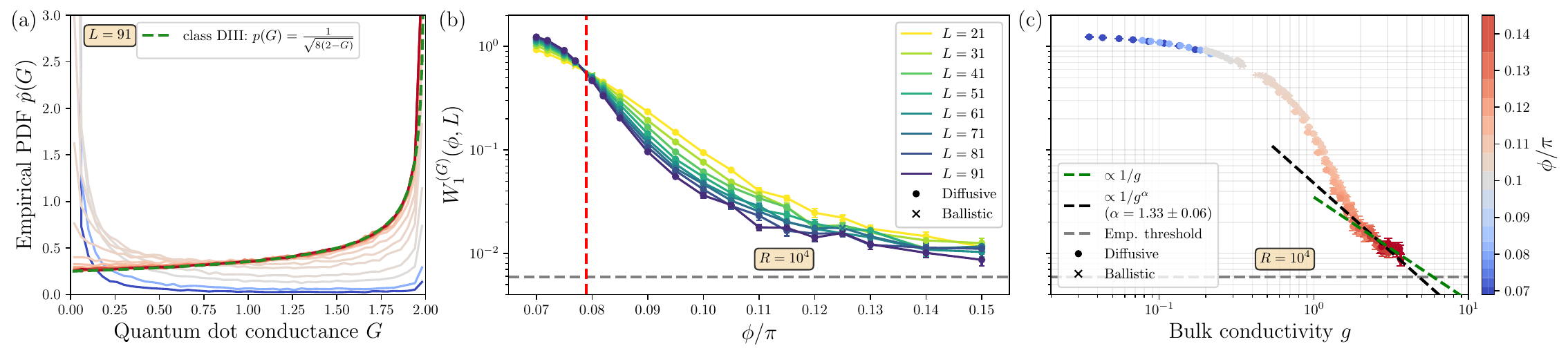}
    \caption{
    Emergent chaotic quantum dot and projected logical ensemble of the honeycomb surface code under partial Pauli twirl. Data averaged over $10^5$ to $10^6$ syndrome realizations for length and width $21\leq L = M \leq 91$. Error bars show SE.
    (a): Empirical probability density function $\hat{p}(G)$ of the quantum dot's conductance $G$ for $L=91$. Above the error threshold $\phi > \phi_{\rm th} \approx 0.079\pi$ [vertical dashed red line in (b)], the empirical PDF approaches the random matrix theory prediction given by a dot made of a Majorana metal in symmetry class DIII (dashed green line). 
    (b): Wasserstein-1 distance $W^{(G)}_1(\phi,L)$ vs $\phi$ between quantum dot's and RMT distributions, or equivalently, the distributions of $G=2\sin^2\theta$ in the projected logical and Haar-hemisphere ensembles. Parameters $(L,\phi)$ that allow for a diffusive (ballistic) metal are shown as circles (crosses). $W^{(G)}_1(\phi,L)$ seems to drop (systematically for the diffusive metal) from a finite value for $\phi < \phi_{\rm th}$ towards a value set by the number $R=10^4$ of syndromes per empirical CDF for $\phi>\phi_{\rm th}$ [horizontal dashed gray line in (b) and (c)]. 
    (c): $W^{(G)}_1(\phi,L)$ vs bulk conductivity $g$ of the network model. $W^{(G)}_1(\phi,L)$ seems to follow a one-parameter scaling $W^{(G)}_1(g)$. In the metallic phase, $W^{(G)}_1(\phi,L)$ decays algebraically with $g$; the dashed green (black) line shows a linear (algebraic) fit.}
    \label{fig:twirl_qdot}
\end{figure*}
We first extract the QEC threshold of the honeycomb surface code under partial Pauli twirl using both the logical error rate $P_{\rm L}$ and bulk conductivity $g$ of the associated scattering network; see Sec.~\ref{sec:HSCthreshold} for the analog discussion for coherent rotations. Using the complex partition functions $\calZ'$ [Eq.~\eqref{eq:partfnTwirl}], we can numerically estimate 
\begin{align}
    P_{\rm L} = \sum_s \min_q \frac{\left| \calZ' \right|^2}{\left| {\cal Z}'_{0,s} \right|^2 + \left| {\cal Z}'_{1,s} \right|^2}
\end{align}by sampling $\eta^{(q)}_{vv'}$ accordingly. We numerically find that $P_{\rm L}$ (not) decreases with the code distance $d\propto L$ below (above) an error threshold $\phi_{\rm th} = (0.079 \pm 0.001)\pi$ [Fig.~\ref{fig:twirl_err_rate}], distinct from the QEC threshold from Sec.~\ref{sec:HSCthreshold}. Using the bulk conductivity $g$, our data is consistent with an insulator-to-metal transition at a critical angle $\phi_{\rm th} = (0.079 \pm 0.002)\pi$ compatible with the observed threshold $\phi_{\rm th}$ [Fig.~\ref{fig:twirl_bulk_g}]. As in the Born-rule case [Fig.~\ref{fig:bulk_g}], $g$ scales with $L/\ell(\phi)$ for some length $\ell(\phi)$: for $\phi<\phi_{\rm th}$, $g(L/\ell(\phi))$ monotonically decreases, signaling that the network model localizes (insulator); conversely for $\phi>\phi_{\rm th}$, $g(L/\ell(\phi))$ increases as $g\propto \pi^{-1} \ln[L/\ell(\phi)]$, which is consistent with a metal in symmetry class DIII~\cite{AndersonTransRMP}; however, the scaling of the conductivity in the metallic phase seems to follow the prediction of a class DIII NL$\sigma$M in the replica limit $N\to 0$ [Fig.~\ref{fig:twirl_bulk_g}] (see also~\cite{yan2026nonlinearsigmamodelsurface, Yang:2026ktu}). This is distinct from the Born-rule case, where $g$ seems to follow the expected $N\to 1$ replica limit in the diffusive regime [cf. Fig.~\ref{fig:bulk_g}].

We now turn to the conductance $G$ of the quantum dot. The analysis mirrors that from Sec.~\ref{sec:qdotNum}. As in the Born-rule case, we find that, in the bulk metallic phase $\phi>\phi_{\rm th}$, the data for the quantum dot agrees remarkably well with the RMT prediction [Fig.~\ref{fig:twirl_qdot}]. The empirical distribution $\hat{p}(G)$ approaches the PDF $\tilde{p}(G)$ predicted by RMT remarkably well [Fig.~\ref{fig:twirl_qdot}(a)]. Moreover, the Wasserstein-1 distance $W_1^{(G)}(\hat{p},\tilde{p})$ between the distribution of $G$ in the quantum dot and from RMT seems to decay with $g$ in the metallic phase [Figs.~\ref{fig:twirl_qdot}(b) and (c)]. In fact, for the largest numerically-accessible $g$, we find that these decay algebraically $\propto g^{-\alpha}$ as in the Born-rule case; however, the exponent $\alpha$ seems to be different.

\section{Symmetry class D numerics}
\label{app:classD}

\begin{figure*}[htp]
    \centering
    \includegraphics[width=\linewidth]{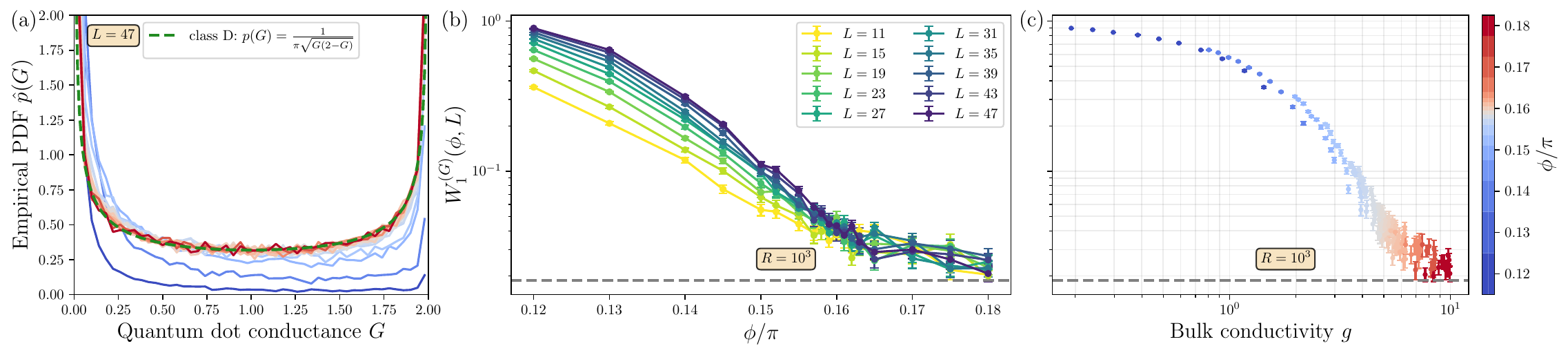}
    \caption{
    Emergent chaotic quantum dot and projected logical ensemble of the triangular surface code under coherent rotations. Data averaged over $2\times10^4$ syndrome realizations for length $11 \leq L \leq 47$ and width $M = 2L-1$. Error bars show SE.
    (a): Empirical probability density function $\hat{p}(G)$ %
    of the quantum dot's conductance $G$ for $L=47$. For $\phi > \phi_{\rm th} \approx 0.16\pi$, the empirical PDF approaches the random matrix theory prediction given by a dot made of a Majorana metal in symmetry class D (dashed green line). 
    (b): Wasserstein-1 distance $W^{(G)}_1(\phi,L)$ vs $\phi$ between quantum dot's and RMT distributions, or equivalently, the distributions of $G=2\sin^2\theta$ in the projected logical and Haar-hemisphere ensembles. $W^{(G)}_1(\phi,L)$ drops from a finite value for $\phi < \phi_{\rm th}$ towards a value set by the number $R=10^2$ of syndromes per empirical CDF for $\phi>\phi_{\rm th}$ [horizontal dashed gray line in (b) and (c)].
    (c): $W^{(G)}_1(\phi,L)$ vs bulk conductivity $g$ of the network model. For $\phi > \phi_{\rm th}$ and large $g$, $W^{(G)}_1(g)$ seems to approach the empirical threshold.
    }
    \label{fig:triangular}
\end{figure*}
In this Appendix, we discuss numerical results for a surface code in symmetry class D with Born-sampled disorder, in addition to the class DIII case discussed in Sec.~\ref{sec:BornNumerics}. Here, we focus on the behavior of the associated class D quantum dot and the projected logical ensemble.

For the simulations, we focus on the triangular-lattice SC with length and width $L = 2M-1$, rotation angle $\phi$, and initial state $\kpsi = \ket{\overline{0}}$ [Sec.~\ref{sec:SCsetup}, Fig.~\ref{fig:overview}(a)]. We use the same methods based on Gaussian (1+1)D circuits as for the honeycomb SC [Sec.~\ref{sec:GaussianCirc}, App.~\ref{app:flo}]. We choose the code distance $d_X$ to be odd such that the final states take the form $\kpsis = \cos(\theta_s/2) \ket{\overline{0}} + e^{i\varphi_s}\sin(\theta_s/2) \ket{\overline{1}}$ with $\varphi_s \in \{ \pi/2, 3\pi/2\}$ and $\theta_s \in [0, \pi/2]$, cf. App.~\ref{app:constraints}, and we numerically reconstruct $\theta_s$ and $\varphi_s$. Using these and the corresponding network model, we estimate the bulk conductivity $g$ and the syndrome-dependent conductance $G_s$ of the quantum dot [cf. Sec.~\ref{sec:BornNumerics}].

Our simulations are consistent with a class D chaotic quantum dot for large enough $\phi$ [Fig.~\ref{fig:triangular}(a)]. For two transmission modes, the expected PDF for the dot's conductance in class D is $\tilde{p}(G) = 1/\pi \sqrt{G(2-G)}$, cf. ~\cite{dahlhaus_prb_2010}, leading to the CDF $\tilde{P}(G) = \pi^{-1} \arccos(1-G)$ for $G \in [0,2]$. The data shown in Fig.~\ref{fig:triangular}(a) suggests that the empirical distribution of $G$ approaches this RMT prediction for $\phi \gtrsim 0.16 \pi$.\footnote{This is consistent with $\xi_{\rm loc} \gg L$, i.e., a metallic grain for the triangular-lattice SC in this regime.} Corroborated with the results in Sec.~\ref{sec:BornNumerics}, this observation suggests that RMT universality can emerge in SC setups for both classes D and DIII.

To probe this approach more thoroughly and its relation with the statistical properties of the PLE, we now turn to the Wasserstein-1 distance $W^{(G)}_1$ between the quantum dot's and the RMT probability distributions, $p$ and $\tilde{p}$, respectively [App.~\ref{app:LErandomness}]. Note that this distribution $\tilde{p}(G)$ with $\vartheta \coloneqq G_s = 2 \sin^2 \theta_s$ is identical to the distribution $p_{\rm H}(G)$ of $\vartheta(\kpsi)$ for $\kpsi$ distributed uniformly on the Bloch sphere subspace spanned by $\varphi_s \in \{ \pi/2, 3\pi/2\}$ and $\theta_s \in [0, \pi/2]$.
We observe that $W^{(G)}_1(\phi,L)$ increases with $L$ for $\phi \lesssim 0.16\pi$, whereas for larger $\phi$, it seems to approach a value set by the number of syndromes $R$ per empirical CDF [Fig.~\ref{fig:triangular}(b)]. As in Sec.~\ref{sec:qdotNum}, the saturation value is $W^{(G)}_1(\hat{p}_{\rm H}, p_{\rm H})$ [shown as the horizontal dashed gray line in Fig.~\ref{fig:triangular}(b) and (c)], where the empirical $\hat{p}_{\rm H}$ is built by sampling $R$ times from $p_{\rm H}(G)$. This observation further suggests that the quantum dot approaches a chaotic regime for large $\phi$, and that, in this chaotic regime, the PLE approaches a class-D-constrained Haar ensemble. While this approach seems to continue with increasing bulk conductivity [cf. Fig.~\ref{fig:triangular}(c)], we note that a considerably larger number of sampled syndromes would be necessary to resolve the precise scaling of $W^{(G)}_1(L, \phi)$ for large $\phi$ and $g$.

\end{document}